\documentclass[debug]{rmaa}

\usepackage{paralist}

\usepackage{psfrag,color}

\usepackage[latin1]{inputenc}

\newcommand{\kms}{km s$^{-1}$}

\newcommand{\hii}{H{\sc ii}}
\newcommand{\msun}{M$_\odot$}

\newcommand{\msunyr}{M$_\odot$ yr$^{-1}$}
\newcommand{\cmdos}{cm$^{-2}$}
\newcommand{\cm}{cm$^{-3}$}
\newcommand{\x}{$\times$}

\newcommand{\radec}{RA, Dec.(J2000)}
\newcommand{\mic}{$\mu$m}
\newcommand{\ca}{$Clump\,A$}
\newcommand{\cb}{$Clump\,B$}
\newcommand{\co}{$^{12}$CO(2-1)}
\newcommand{\tco}{$^{13}$CO(2-1)}
\newcommand{\hcop}{HCO$^{+}$(1-0)}
\newcommand{\hnc}{HNC(1-0)}
\newcommand{\ndhp}{N$_{2}$H$^{+}$(1-0)}
\newcommand{\cdo}{C$^{18}$O(2-1)}
\newcommand{\irdc}{IRDC SDC\,341.232-0.268}
\newcommand{\sdc}{SDC\,341.232-0.268}
\newcommand{\egoa}{G\,341.24-0.27}
\newcommand{\egob}{G\,341.23-0.27}
\newcommand{\egoc}{G\,341.22-0.26(a)}
\newcommand{\egod}{G\,341.22-0.26(b)}  

\title{Millimeter and Far-IR study of the \irdc} 

\author{
  M. M. Vazzano,\altaffilmark{1} 
  C. E. Cappa,\altaffilmark{2}
  V. Firpo, \altaffilmark{4,5}
  C. H. L\'opez-Caraballo, \altaffilmark{6}
  M. Rubio, \altaffilmark{3}
  and N. U. Duronea, \altaffilmark{1}
  }

\altaffiltext{1}{Instituto Argentino de Radioastronom{\'{\i}}a, CONICET, CCT-La Plata, Villa Elisa, Argentina.}
\altaffiltext{2}{Facultad de Ciencias Astron\'omicas y Geof{\'{\i}}sicas, Universidad Nacional de La Plata, La Plata,  Argentina.}
\altaffiltext{3}{Departamento de Astronom{\'{\i}}a, Universidad de Chile, Chile.}
\altaffiltext{4}{Gemini Observatory, Southern Operations Center, La Serena, Chile.}
\altaffiltext{5}{Departamento de Astronom{\'{\i}}a y F{\'{\i}}sica, Universidad de la Serena, La Serena, Chile.}
\altaffiltext{6}{Departamento de Matem\'aticas, Universidad de La Serena, La Serena, Chile.}

\shortauthor{M. M. Vazzano et al.}
\shorttitle{RevMexAA Main Journal Demo Document}

\fulladdresses{
\item Vazzano M. M. and Duronea N. U.: Consejo Nacional de Investigaciones Cient{\'{\i}}ficas y T\'ecnicas, Argentina, Instituto Argentino de
Radioastronom{\'{\i}}a, CONICET, CCT La Plata, C.C.5, 1894, Villa Elisa, Argentina (mvazzano@iar.unlp.edu.ar, duronea@iar.unlp.edu.ar).
\item Cappa C. E.: acultad de Ciencias Astron\'omicas y Geof{\'{\i}}sicas, Universidad Nacional de la Plata, Paseo del Bosque s/n, 1900, La Plata, 
Argentina (ccappa@fcaglp.unlp.edu.ar).
\item Firpo V: Gemini Observatory, Southern Operations Center, C/o AURA, Casilla 603, La Serena, Chile (vero.firpo@gmail.com)
\item L\'opez Caraballo C. H.: Departamento de Matem\'aticas, Universidad de La Serena, Av. Juan Cisternas 1200, La Serena, Chile (lopezcaraballoch@gmail.com).
\item Rubio M.: Departamento de Astronom{\'{\i}}a, Universidad de Chile, Casilla 36, Santiago de Chile, Chile (mrubio@das.uchile.cl).
}

\listofauthors{M. M. Vazzano, C. E. Cappa, M. Rubio, V. Firpo, C. H. L\'opez-Caraballo, \& N. U. Duronea}
\indexauthor{Vazzano, M. M.}
\indexauthor{Cappa, C. E.}
\indexauthor{Rubio, M.}
\indexauthor{Firpo, V.}
\indexauthor{L\'opez-Caraballo, C. H.}
\indexauthor{Duronea, N. U.}

\abstract{We analyze the molecular gas and dust associated with the infrared dark cloud \sdc\ in order to investigate the characteristics and parameters of the gas, 
determine the evolutionary status of four embedded EGO candidates and establish possible infall or outflow gas motions.   
   We based our study on \co, \tco, and \cdo\ data obtained with the APEX telescope, molecular data of high density tracers from the MALT90 survey and IR images from   
   {\it Spitzer}, {\it Herschel} and ATLASGAL.    
   The study reveals two clumps at $-$44 \kms\ towards the IRDC, with densities of $>$ 10$^4$\cm, typical of IRDCs, while high density tracers show H$_2$ densities $>$ 10$^{5}$.
    FIR images reveals the presence of cold dust linked to the molecular clumps and EGOs. A comparison of spectra of the optically thin and optically thick molecular lines towards the EGOs  suggests the existence of infall and outflow motions.}

\resumen{
Analizamos el gas molecular y el polvo asociado a la nube oscura infrarroja  \sdc\ con el fin de investigar las características y parámetros físicos del gas, 
determinar el estado evolutivo de los cuatro EGOs embebidos  y establecer posibles movimientos de acreción o flujo molecular.   
Nos basamos en datos de \co, \tco\ y \cdo\ obtenidos con el telescopio APEX, trazadores de alta densidad extraídos de MALT90, e 
imágenes infrarrojas de  {\it Spitzer}, {\it Herschel} y ATLASGAL. 
El estudio revela dos grumos moleculares a $-$44 \kms\ coincidentes con la IRDC con una densidad  $>$ 10$^4$\cm, típica de IRDCs.
Los trazadores de alta densidad arrojan densidades de H$_2$ $>$ 10$^5$. 
Las imágenes en el lejano IR muestran polvo frío asociado a los grumos moleculares y a los EGOs. La comparación de espectros moleculares ópticamente gruesos y finos sugiere 
la existencia de acreción y flujos moleculares.
}

\addkeyword{ISM: kinematics and dynamics}
\addkeyword{ISM: molecules}
\addkeyword{Stars: formation}
\addkeyword{(ISM:) dust, extinction}
\addkeyword{ISM: clouds}
\addkeyword{ISM: individual objects: \irdc}

\begin{document}
\maketitle

\section{Introduction}
\label{sec:intro}

Infrared Dark Clouds (IRDCs) are sites of recently star formation within molecular clouds. 
These are detected as dark silhouettes against a bright background in the mid-infrared.
These regions, which are detected in the FIR, are believed to be places that may contain compact cores which probably host the early stages of high-mass star formation
(e.g., \citealt{rathborne2007,chambers2009}). 
 In fact, their cold (T $<$ 25\,K) and dense (10$^2$-10$^4$\,\cm) ambient conditions, and their sizes of $\sim$1-3\,pc and masses of 10$^2$-10$^4$\,\msun\ 
 \citep{rathborne2006, rathborne2007} favor star formation mechanisms.  Evidences of active high-mass star formation in IRDCs are inferred from the
 presence of ultracompact \hii\ regions \citep{battersby2010},  hot cosres \citep{rathborne2008}, embedded 24\,$\mu$m sources \citep{chambers2009},
 maser emission \citep{Wang2006,chambers2009}, and/or outflow and infall processes (e.g. \citealt{sanhueza2010, chen2011, ren2011, beuther2002, Zhang2007}). 

   \begin{figure*}
   \centering
   \includegraphics[angle=0,width=12cm]{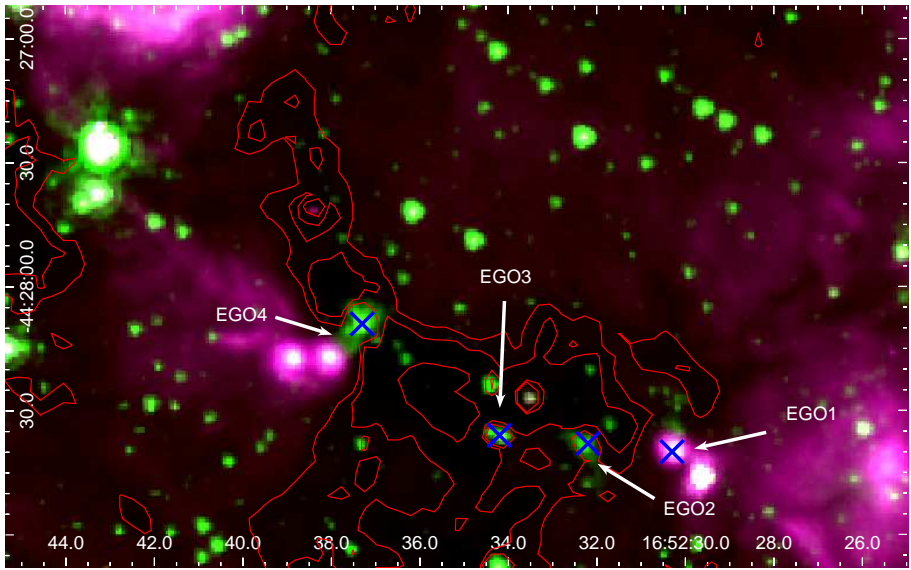}
    \caption{Composite image showing the {\it Spitzer}-IRAC emission at  4.5 (green) and 8.0\,$\mu$m (red), and  {\it Spitzer}-MIPS emission  at
    24\,$\mu$m (blue) of the region of \sdc. Contours correspond to the emission at 8.0\,$\mu$m at 80, 90, and 95 MJy/ster. The positions of the
    candidate EGOs are marked with blue crosses. }
   \label{Fig1}
\end{figure*}

 Many IRDCs are also associated with extended enhanced 4.5 \mic\ emission (named Extended Green Objects, EGOs, for the common coding of the [4.5] band in
 green in the 3-color composite IRAC images), the  so-called ``green fuzzies'', indicating the presence of shocks. Following photometric criteria, \citet{cyganowski2008} suggested that
 most of EGOs fall in the region of the color-color space  occupied by the youngest MYSOs, surrounded by accreting envelopes (see Figure 13 in their work).
 This hypothesis is supported by \citet{Chenetal2010}, whose observations are consistent with the speculation that EGOs trace a population with ongoing outflow activity and active
rapid accretion stage of massive protostellar evolution. 
 The emission at 4.5\,$\mu$m includes both H$_2$ ({\rm v} = 0-0, S(9, 10, 11)) lines and CO ({\rm v} = 1-0) band heads 
\citep{cyganowski2008}, which are excited by shocks, such as those expected when protostellar outflows impinge on the ambient interstellar medium (ISM)  \citep{cyganowski2008}.
   
 Using IRAC images, \citet*{peretto2009} catalogued and characterized many IRDCs using 8 and 24\,$\mu$m {\it Spitzer} data.
 In this  catalogue, the IRDCs were defined as connected structures with H$_2$ column density peaks $N_{\rm H_2}  >$  2$\times$10$^{22}$\,\cmdos\
 and boundaries defined by  $N_{\rm H_2}$ = 1$\times$10$^{22}$\,\cmdos. Single-peaked structures found in the H$_2$ column density maps were identified by the authors as fragments.
 These dark and dense patches are associated with molecular gas and dust emission \citep{sanhueza2012}.

 As a part of a project aimed at studying in detail the physical properties of the IR bubble complex S21-S24, we present here an analysis of the molecular gas 
 and dust associated with SDC\,341.232-0.268, a poorly studied IRDC of the southern hemisphere, which is member of this complex.
  Our aims are to determine the physical parameters and kinematics of the gas and to chracterize the interstellar dust associated with the IRDC, to investigate the 
  presence of embedded young stellar objects and to  establish possible infall or   outflow gas motions. 
 To accomplish this project, we analyze molecular data of the \co, \tco, and \cdo\ lines obtained with the APEX telescope, and \hcop, \hnc, and \ndhp\
 lines from  the MALT90 survey \citep{Jackson13, Foster13, Foster11}\footnote{http://malt90.bu.edu/data.html}, mid- and far-IR continuum data from 
 {\it {\it Spitzer}}-IRAC, {\it Herschel}-PACS and -SPIRE, and ATLASGAL (APEX) images \citep{schuller2009}. 
 We introduce the \irdc\ in \S ~\ref{anteced}, the used database in \S ~\ref{data}, the results of the molecular gas in \S ~\ref{results-mol}, 
 and those of the infrared emission in \S~\ref{dust}, signs of star formation are discussed in \S~\ref{dis}, and a summary in \S~\ref{summary}.\\

 \section{\irdc}\label{anteced}
  
 The \irdc\ (\radec= \hbox{16:52:35.67}, \hbox{--44:28:21.8)} has a mid-IR radiation field intensity $I_{\rm MIR} \sim$ 88.3 MJy  ster$^{-1}$ at 8.0\,$\mu$m 
and a peak opacity of 1.09. From the Multiband Infrared Photometer for {\it Spitzer} (MIPS) images \citet*{peretto2009} detected a 24\,$\mu$m point source in the 
field of the IRDC and two clumps. The clumps were classified by \citet*{bergin2007a} as structures with sizes $\sim$10$^{-1}$-100 pc, masses $\sim$10-10$^{3}$ M$_{\odot}$, 
and volume densities $\sim$10$^3$-10$^4$ \cm.

Figure~\ref{Fig1} shows the {\it Spitzer}-IRAC images at 4.5 (green) and 8.0\,$\mu$m (red), and the {\it Spitzer}-MIPS image at 24\,$\mu$m (blue) in the region of \sdc. 
Green contours correspond to low emission levels at 8\,$\mu$m, revealing a region of high extinction at 8.0\,$\mu$m at the location of the IRDC, in contrast with the environment, 
a typical feature of these objects. The 90 MJy/ster contour delineates the IRDC. 
Four candidate EGOs catalogued by  \citet{cyganowski2008} appear projected toward this region: the ``likely'' MYSO outflow candidates \egod (R.A., Decl.(J2000) = 16:52:30.3, --44:28:40.0), \egoc\ (16:52:32.2, --44:28:38.0), 
the ``possible'' MYSO outflow candidate \egob\ (16:52:34.2, --44:28:36.0),  and  the ``likely'' MYSO outflow candidate \egoa\ (16:52:37.3, --44:28:09.0). 
These authors categorized the EGOs as a ``likely'' or ``possible'' MYSO outflow candidate based primarily on the angular extent of the extended excess 4.5 \mic\ emission.
Any  source  in  which  it  was  possible that multiple nearby point sources and/or image artifacts from a bright  IRAC source  could  be  confused  with truly extended 4.5 \mic\ 
emission was considered a ``possible''  candidate, which are likely still good YSO candidates, but not necessarily MYSOs with outflows and so likely to be actively accreting. 

From here on, they will be named EGO\,1, EGO\,2, EGO\,3, and EGO\,4,
in increasing order of R.A. EGO\,1 is associated with 8 and 24\,\mic\ emission and is  located  in the border of the IRDC,
while EGOs 2, 3, and 4 are detected at 4.5 and 8\,$\mu$m. 
EGO\,1 and EGO\,4 coincide with ATLASGAL Compact Sources AGAL\,341.219-00.259 and AGAL\,341.236-00.271. Their integrated flux densities are
12.85$\pm$2.20 Jy and 16.81$\pm$2.80 Jy, and their effective radii are $\sim$ 37\arcsec\ and 47\arcsec, respectively \citep{contreras2013}. 

Methanol maser emission was detected towards EGO\,2 and EGO\,4 at 6 and 95 GHz, within the velocity range $-$43 to $-$52\,\kms \citep{caswell2010,chen2011,hou2014,yang2017}. 
Methanol masers provide signpost to the very earliest stages of the massive star formation process, prior the onset of the UCHII region phase.
They are associated with embedded sources whose bolometric luminosities suggest the will soon become OB stars \citep{Burtonetal2002,Sobolevetal2005}.
The masers are independent tracers and they give proven signatures of ongoing star formation as the methanol masers (see \citealt{Ellingsen2006}; \citealt{Breenetal2013}). 
In other hand, masers of water or hydroxyl have also been detected in in star-forming regions, as well as in evolved stars or supernova remnants. Therefore, 
the presence of these maser types in the vicinity of these EGOs could be an excellent signspot of ongoing star formation.

\section{Database}\label{data}

\subsection{Molecular line observations}


\begin{table}
\caption[]{Observational parameters of the molecular transitions.}
\centering
\label{tabla2}
\begin{tabular}{lcccc}
\hline 

Transition & $\nu_0$  & $\theta$  & $\Delta {\rm v}_{res}$ & rms \\
           & GHz    & \arcsec\    &  \kms             &     K   \\
\hline
\co       & 230.538 &  30        &     0.1            &   0.35  \\
\tco      & 220.398 &  28.5      &     0.1            &   0.35 \\
\cdo      & 219.560 &  28.3      &     0.1            &   0.35  \\
HNC(1-0)  & 90.664 &  38.0      &     0.11            &   0.35  \\
HCO$^+$(1-0) & 89.189 &  38.0     &     0.11         &   0.35  \\
N$_2$H$^+$(1--0) & 93.173 &  38.0  &     0.11        &   0.35  \\
\hline 
\end{tabular}
\end{table}


   The \co, $^{13}$CO(2-1), and C$^{18}$O(2-1) data were acquired with the APEX-1 receiver of the Swedish Heterodyne Facility Instrument (SHeFI;  \citealt{vassilev2008}) in the Atacama Pathfinder EXperiment (APEX) telescope, located in the Puna de Atacama (Chile). The backend for the observations was the eXtended bandwidth Fast Fourier Transform Spectrometer2 (XFFTS2) with a 2.5 GHz bandwidth divided into 32768 channels.
   The main parameters of the molecular transitions (rest frequency $\nu_0$, half-power beam-width $\theta$, velocity resolution  $\Delta {\rm v}_{\rm res}$, and {\it rms} noise of the individual spectra obtained using the OTF  mode) are listed in Table~\ref{tabla2}.  The selected off-source position free of molecular emission was \radec = (16:36:40.56, $-$42:03:40.6).\

   Calibration was done using the chopper-wheel technique. The antenna temperature scale was converted to the main-beam brightness-temperature scale  by $T_{\rm mb}$ = $T_{\rm A}$/$\eta _{\rm mb}$, where $\eta _{\rm mb}$ is the main beam efficiency. For the SHeFI/APEX-1 receiver we adopted $\eta _{\rm mb}$ = 0.75. 
   Ambient conditions were good, with a precipitable water vapuor (PWV) between 1.5 $-$ 2.0 mm.\

 The molecular spectra were reduced using the CLASS90 software of the IRAM's GILDAS software package.
 
  In addition, we used molecular data from the Millimetre Astronomy Legacy Team 90 GHz Survey (MALT90) taken with the Mopra spectrometer (MOPS). 
We used beam efficiencies between 0.49 at 86 GHz and 0.42 at 230\,GHz \citep{Ladd2005}. The data analysis was conducted with CLASS90 software.  Emission was detected in the HCO$^+$(1-0), HNC(1-0), and N$_2$H$^+$(1-0) lines, which were used to detect high density regions within \sdc. Their main parameters are included in  Table~\ref{tabla2}.


\subsection{Images in the infrared}

We used near- and mid-infrared (NIR, MIR) images from the {\it {\it Spitzer}}-IRAC archive at 4.5 and 8.0\,$\mu$m of the Galactic Legacy Infrared Mid-Plane Survey 
Extraordinaire (GLIMPSE) \footnote{http://irsa.ipac.caltech.edu/data/SPITZER/GLIMPSE/} \citep{benjamin03}, and the Multiband Imaging Photometer for {\it {\it Spitzer}} 
(MIPS) image at 24\,$\mu$m from the MIPS Inner Galactic Plane Survey (MIPSGAL)\footnote{http://irsa.ipac.caltech.edu/data/SPITZER/MIPSGAL/} \citep{carey2005} to delineate
the IRDC and investigate their correlation with the EGO candidates.

To trace the cold dust emission we utilized far-infrared ({\sc FIR}) images from the {\em Herschel Space Observatory} belonging to the Infrared GALactic (Hi-GAL) plane 
survey key program \citep{molinari2010}. The data were carried out in parallel mode with the instruments PACS \citep{Poglitsch2010} at 70 and 160\,$\mu$m, and SPIRE 
\citep{Griffin2010} at 250, 350, and 500\,$\mu$m. The angular resolutions for the five photometric bands spans from 8\arcsec\ to 35\farcs 2 for 70\,$\mu$m to 500\,$\mu$m.
{\em Herschel} Interactive Processing Environment (HIPE v12\footnote{{\em HIPE} is a joint development by the Herschel Science Ground 
Segment Consortium, consisting of ESA, the NASA Herschel Science Center, and the HIFI, PACS and SPIRE consortia members, see http://herschel.esac.esa.int/HerschelPeople.shtml.},
\citealt{Ott2010}) was used to reduce the data, with reduction scripts from standard processing. 
The data reduction and calibration (including zero-level and color correction) is described in details in section\,2.2 of \citet{cappa2016}.

We also used images at 870\,$\mu$m from the APEX Telescope Large Area Survey of the Galaxy (ATLASGAL) \citep{schuller2009} with a beam size of 19\farcs 2. This survey 
covers the inner Galactic plane, with an {\it rms} noise in the range 0.05--0.07 Jy beam$^{-1}$. The calibration uncertainty in the final maps is about 15\%.

 
 \section{Molecular characterization of the IRDC}\label{results-mol}
 
 \subsection{CO data:  morphological and kinematical description} \label{mol-kin}
 

\begin{figure*}
   \centering
   \includegraphics[angle=0,width=1.0\textwidth]{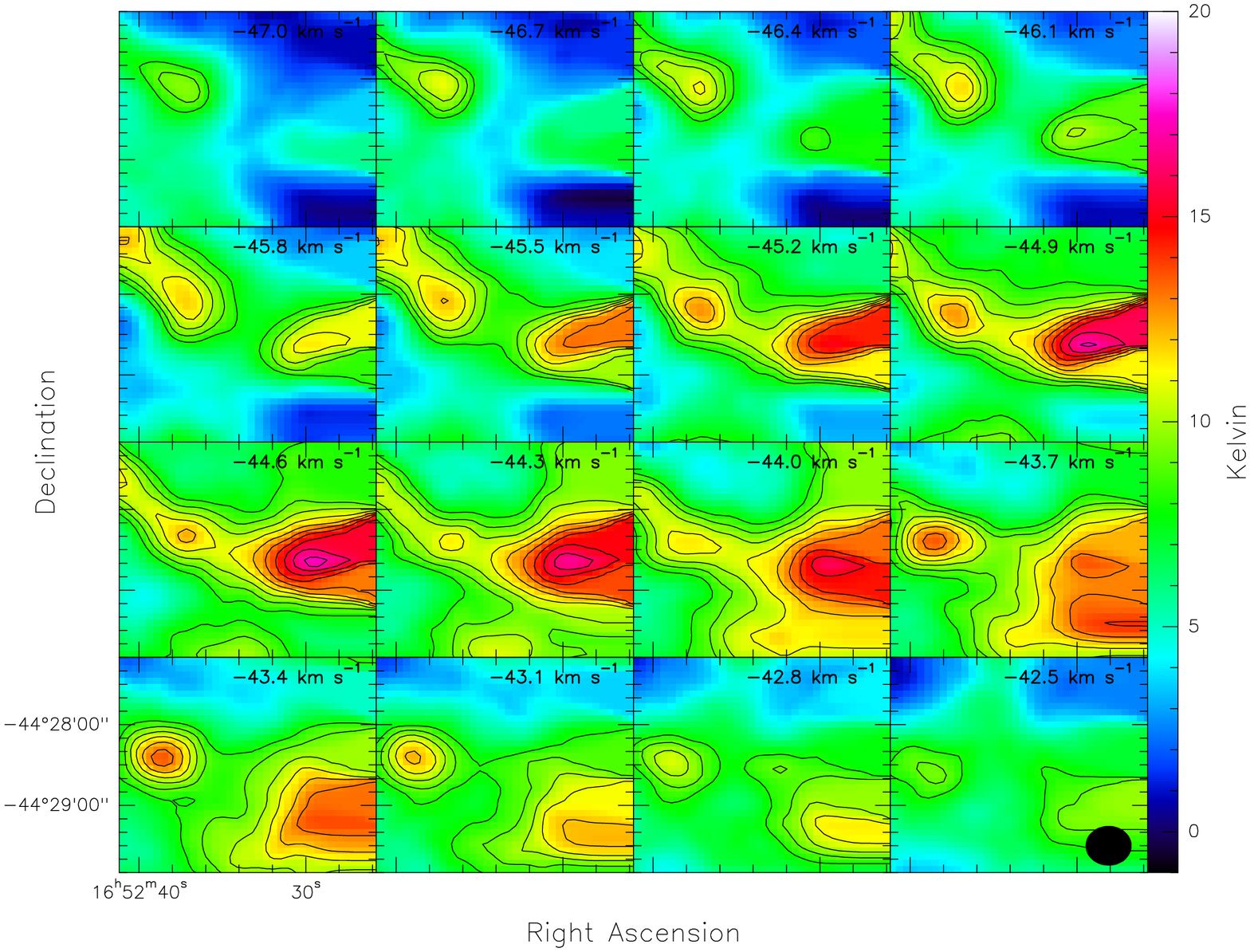}
   \hspace{3mm}
   
    \includegraphics[angle=0,width=1.0\textwidth]{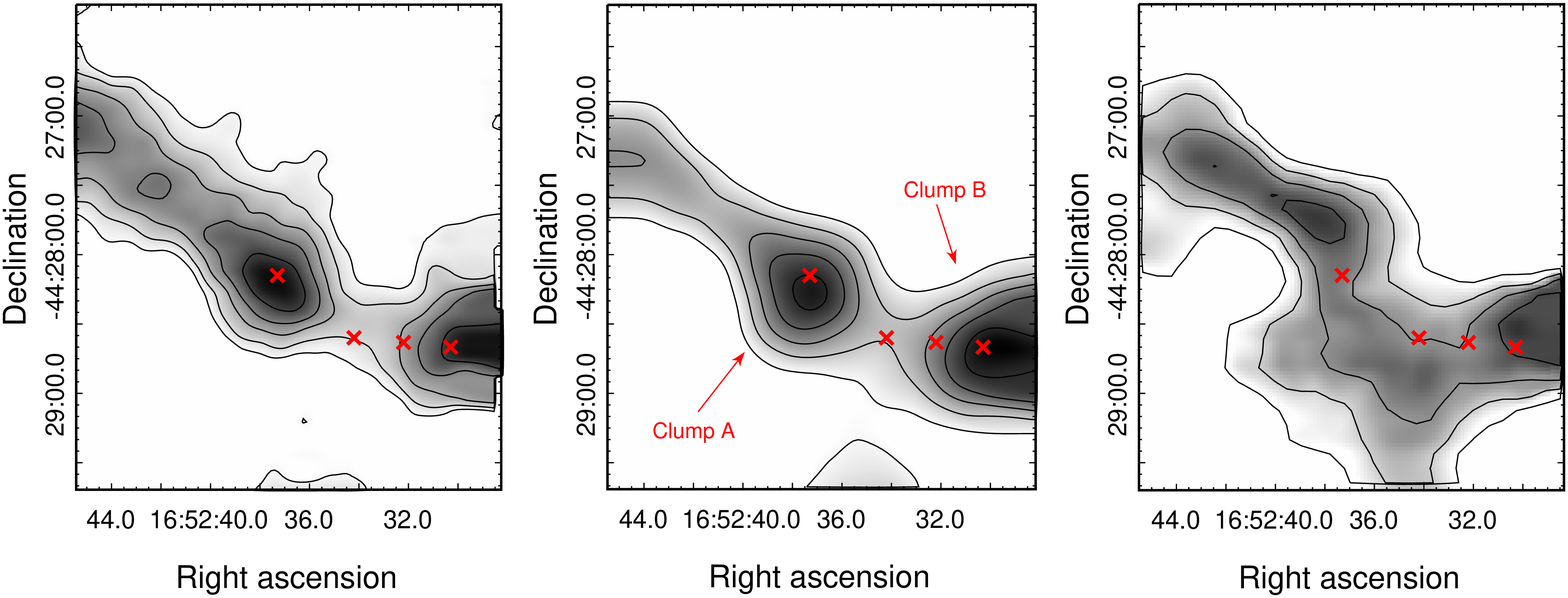}
    \caption{{\it Upper panel:} T$_{\rm mb}$-maps of the \tco\ emission within the velocity interval from --47.0 to --42.5\,\kms\ in steps of 0.3\,\kms. Contours range from 8 to 17\,K in steps of 1\,K in T$_{\rm mb}$. 
{\it Lower left panel:} Average \cdo\ emission in the range --47.2 to --43.8\,\kms. Contours are 0.8, 1.0, 1.3, 1.6, and 2.0\,K. The red crosses mark the position of the EGOs.
{\it Lower central panel:} Average \tco\ emission in the range $-$47.0 to $-$42.8\,\kms. Contours range from 6.2 to 9.4\,K in steps of 0.8\,K.
{\it Lower right panel:} Average \co\ emission in the range $-$48.2 to $-$42.5\,\kms. Contours range from 10 to 13\,K in steps of 1\,K.} 
   \label{prom-13CO}
\end{figure*}

 To visualize the spatial distribution of the molecular emission the upper panel of Fig.~\ref{prom-13CO} exhibits the \tco\ brightness temperature distribution from $\sim$ --47.0 to --42.5 
 \kms\ in steps of 0.3 \kms. The most intense \tco\ emission appears in the range $-$45.2 to $-$44 \kms. 
 The bottom central  panel of Fig.~\ref{prom-13CO} shows the average \tco\ emission 
 in the range $-$47.0 to $-$42.8 \kms, revealing two molecular clumps centered at \radec = (16:52:37.06, $-$44:28:13.9) {\it (Clump A)} and \radec = (16:52:29.72, $-$44:28:40.1)  {\it (Clump B)}. 
 The coincidence of the clumps with EGO\,1, 2, and 4 (indicated with crosses in this figure) is clear. 
 The  effective radii of the clumps are 37\farcs 6 and 45\farcs 6 for \ca\ and \cb, respectively, as obtained taking into account the average \tco\ emission 
 higher that 6.2 K. Finally, the bottom left and right panels of the same figure display the \cdo\ and \co\ average emission in velocity intervals similar to that in the image of \tco. 
 Both clumps A and B are detected at \cdo\ revealing the existence of dense molecular gas in the clumps. The \co\ emission delineates the IRDC but the clumps are not well defined.  
 

\begin{figure*}
   \centering
   \includegraphics[angle=0,width=12cm]{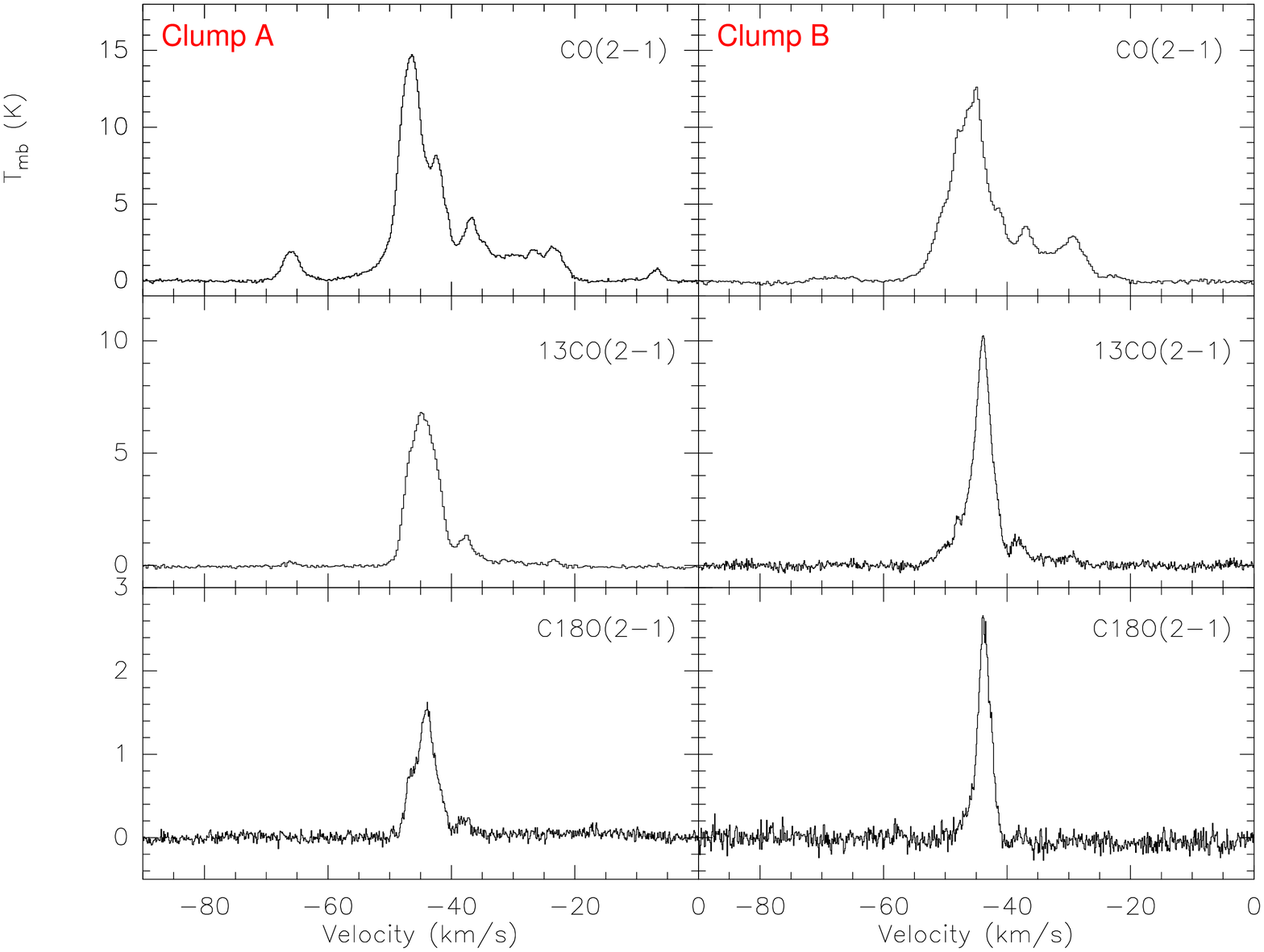}
    \caption{Averaged \co, \tco\ and \cdo\ spectra toward \ca\ (left) and \cb (right).} 
   \label{perfiles}
\end{figure*}

Figure~\ref{perfiles} shows the averaged \co, \tco, and \cdo\ spectra within the area corresponding to \ca\ and \cb. The {\it rms} noise of the  averaged 
spectra is 0.019 K for \ca, and 0.017 for \cb.
  For \ca, the \co\ line exhibits a multi-peak structure with components in the interval from --70 to --20 \kms, with the most intense features between 
  --48 and --35 \kms. Three velocity componentes are detected within the last velocity interval, peaking at --46.6 \kms, --42.2 \kms, and --36.8 \kms.   
  The \tco\ and \cdo\ spectra peak at around --44.0 \kms, coincident with a minimum in the \co\ spectrum. Components outside this range will be analyzed in  \S~\ref{outflows}.
For \cb, the \co\ line also shows a multi-peak structure between --55 and --25 \kms, with the maximum at --45.4 \kms. The \tco\ and  \cdo\ profiles peak at \hbox{--44.0} \kms.
    
Taking into account that the \cdo\ emission is generally optically thin, we adopted systemic velocities (${\rm v}_{\rm sys}$) of --44.0 \kms\ for \ca\ and \cb. 
The adopted systemic velocity coincide with the velocity for EGO\,2 and 4 reported by \citet{yang2017} (--44.6 and --44.9, respectively) from methanol maser emision, 
and from masers by \citet{chen2011}. It is also compatible with the velocity of NH$_3$ clouds identified by \citet{purcell2012} at --43.7 \kms\ at a position distant 15\arcsec\ from EGO\,3, and from the CS(2-1) 
line emission obtained by 
\citet{bronfman1996}, who observed   toward  [\radec = (16:52:34.2, --44:28:36.0)], revealing  the presence of high density regions ($n_{crit}$ = 3.0$\times$10$^5$ \cm).  
 Circular galactic rotation models predict that gas with velocities of --44 \kms\ lies at the near kinematical distance of 3.6 kpc (see, for example, 
 \citealt*{brand1993}). Adopting a velocity dispersion of  6\,\kms, the uncertainty in distance is 0.5\,kpc (15\%). The distance coincides also with that of S\,24 (Cappa et al. 2016) indicating that the IRDC belongs to the same complex.


\subsection{Parameters derived from CO data}\label{COparameters}


\begin{table*}
\centering
\caption{Parameters of the gaussian fits for \ca\ and \cb.}

 \begin{tabular}{lccccc}
\hline
      &  Line  &  $v$   &  $\Delta$v   & T$_{\rm peak}$ & Area \\
      &        &  \kms\ &    \kms\     &    K           & K \kms\ \\
\hline
\ca\  & \tco\  & --44.56(0.03) & 5.56(0.07) & 6.82 & 40.33(0.41) \\
      & \cdo\  & --44.23(0.02) & 4.46(0.05) & 1.37 &  6.49(0.06) \\
\hline
\cb\  & \tco\  & --43.85(0.01) & 2.56(0.08) & 9.52 & 25.96(0.18) \\
      & \cdo\  & --43.70(0.01) & 2.61(0.03) & 2.46 & 6.84(0.07) \\
\hline
 \end{tabular}
\label{tab:fits-lines}
\end{table*}


Bearing in mind that the spatial distribution of the \tco\ and \cdo\ emission coincide with the IRDC, we used these lines to estimate the main parameters of the molecular gas assuming  local thermodynamic equilibrium (LTE). All the calculations were carried out from the gaussian fits to the lines averaged in the area of \ca\ and \cb\  shown in Table \ref{tab:fits-lines}. The columns list the central velocity of each component $v$, the full-width at half-maximum, the peak temperature, and the integrated emission. We apply the following expression to estimate optical depths \citep[e.g.][]{Scoville86}:
\begin{equation}
 \frac{T^{13}_{\rm mb}}{T^{18}_{\rm mb}} = \frac{1 - e^{-\tau^{13}}}{1 - e^{-\tau^{13}/7.6}}
\end{equation}
\noindent where $\tau^{13}$ is the optical depth of the \tco\ gas and 7.6 = [\tco]/[\cdo] \citep{sanhueza2010} is the isotope abundance ratio. The \tco\ optical depths are indicated in column 2 of Table \ref{gas-para}. To estimate the \cdo\ optical depth we use 
\begin{equation}
 \tau_{18} = \frac{1}{7.6} \hspace{1mm} \tau_{13} \left( \frac{\Delta {\rm v}_{13}}{\Delta {\rm v}_{18}} \right) \left( \frac{\nu^{13}}{\nu^{18}} \right) ^2
\end{equation}
\noindent where $\Delta {\rm v}$ is the full-width at half-maximum of the \cdo\ and \tco\ profiles.
The results are listed in column 3 of Table \ref{gas-para}. 
Bearing in mind that \tco\ is moderately optically thick we calculate the excitation temperature from the \tco\ line using
\begin{equation}
T_{\rm exc} = \frac{T^*} {ln\left[ \left( \frac{T_{\rm mb}^{13}}{T^{*}}  +  \frac{1}{ { {\rm exp}(T^{*}/T_{\rm bg})-1} )} \right)^{-1} + 1 \right]},
\label{texc}                       
\end{equation}
\noindent where $T^*$ = $\frac{h \nu}{k}$, $\nu$ is the frequency of the \tco\ line and $T_{\rm bg}$ = 2.7 K. Values estimated for $T_{\rm exc}$  are 
listed in Table~\ref{gas-para}. 

\begin{table*}
 \begin{changemargin}{-5cm}{-5cm}
\centering
\caption[]{Derived parameters of the molecular components.}
\label{gas-para}
\begin{tabular}{lcccccccccccc}
\hline
        &  $\tau_{13}$ & $\tau_{18}$ & T$_{\rm exc}$ &  $N({\rm ^{13}CO})$ & $N({\rm H_2})$ & $r$  & $M({\rm H_2})$ & $n_{{\rm H2}}$ & $M_{\rm vir}$ & $\frac{M_{\rm vir}}{M({\rm H_2})}$  \\
        &             &            &     K       & 10$^{16}$ cm$^{-2}$  & 10$^{22}$ cm$^{-2}$ & pc & $M_{\odot}$ & 10$^4$cm$^{-3}$  &   $M_{\odot}$ &                               \\
\hline
\ca     & 3.4 & 0.6 &  11.5 &  14.20  &  7.1  &  0.7 &  2400$\pm$960  & 2.6$\pm$1.3  & 1000-1670 & 0.41-0.69    \\
\cb     & 1.1 & 0.1 &  14.4 &  4.12  &  2.06  &  0.9 &  1200$\pm$480  & 0.5$\pm$0.2  & 750-1240 & 0.62-1.0    \\
\hline
\end{tabular} 
  \end{changemargin}

\end{table*}

The column density for \tco\ was derived using \citet{RohlfsandWilson2004}
\begin{eqnarray}
 N(^{13}{\rm CO}) = 2.4 \times 10^{14} {\rm exp}\left(\frac{T^*}{T_{\rm exc}}\right) \times \frac{T_{exc}}{1-e^{-T^*/T_{\rm exc}}}\int \tau_{13} d{\rm v}
\end{eqnarray}
In this case we used the approximation for $\tau_{13}$ $>$ 1, 
\begin{eqnarray}
T_{exc} \int \tau_{13} d{\rm v} \simeq \times \frac{\tau^{13}}{1-exp(-\tau^{13})}  \int T_{\rm mb}^{13} d{\rm v}
\end{eqnarray}
This approximation helps to elliminate to some extent optical depths effects. The integral was evaluated  as $T_{mean-mb} \times \Delta {\rm v}$,  where $T_{mean-mb}$ is equal to  the average $T_{mb}$ within the  area of each clump.

To estimate $N(\rm H_2)$ (column 6 in Table \ref{gas-para}) we adopted an abundance [H$_2$][$^{13}$CO]= 5$\times$10$^5$ \citep*{dickman1978}.
Then, the molecular mass (column 9 in Table \ref{gas-para}) is calculated from the equation
\begin{equation}
M(H_2) = (m_{sun})^{-1} \mu\ m_{\rm H } \ A \ N(H_2) \ d^2
\label{eq:masa}
\end{equation}

\noindent where $m_{sun}$ = 2$\times$10$^{33}$\,g is the solar mass, the mean molecular weight $\mu$ = 2.76 (which includes a relative helium abundance of 25\% by mass, \citealt{allen1973}), and $m_{\rm H}$ is the hydrogen atomic mass. In this expression $d$ is the distance, $N(H_2)$ is the H$_{2}$ column density, and $A$ is the area of the source in cm$^{-2}$. 
The effective radius of each clump as seen in the \tco\ line and the volume densities are listed in colums 7 and 9 in Table \ref{gas-para}. 

The ambient volume densities, $n_{\rm H_2}$, was calculated assuming a spherical geometry for the clumps, using the formula
\begin{equation}
n_{\rm H_2} = \frac{M(H_{\rm 2})}{\frac{4}{3}\pi\ r^3\mu \hspace{0.7mm} m_H}
\end{equation}.
 
Ambient densities are listed in column 9 of Table \ref{gas-para}.
The parameters calculated in this work agree with \citet*{bergin2007a}.

The virial mass can be determined following \citet{macLaren1988}:
\begin{equation}  
\frac{M_{\rm vir}}{M_{\odot}} = k_2 \left[\frac{r}{\rm pc}\right] \left[\frac{\Delta {\rm v}^2}{\rm km s^{-1}}\right]
\end{equation}

\noindent where $r$ and $\Delta$v are the radius of the region and the velocity width measured from the gaussian fit of the \cdo\ emission, and $k_2$ depends on the 
geometry of the ambient gas in the region, being 190 or 126 according to $\rho \propto r^{-1}$ or $\rho \propto r^{-2}$, respectively. $M_{\rm vir}$ values are 
listed in column 10 in Table \ref{gas-para}. 
The ratios $M_{vir}/M(H_2)$ suggest that both clumps may collapse to form new stars, since if the virial mass value is lower than the LTE mass value 
then this clump does not have enough kinetic energy to stop the gravitational collapse.

Uncertainties in both the molecular mass derived using LTE conditions, $M(\rm H_2)$, and the  virial mass $M_{\rm vir}$ are affected by distance indetermination 
(15\%) yielding 30\% error in  $M(\rm H_2)$ and 15\% in $M_{\rm vir}$. Inaccuracies in the borders of the clumps originate errors in their sizes and thus additional 
uncertainties in the masses,  suggesting errors of 40\% in $M(\rm H_2)$. Virial masses are not free of uncertainties, because of the existence of magnetic field
support that may overestimate the derived values \citep[see][]{macLaren1988} and the unknown density profile of the clump.

\subsubsection{Analysis of the MALT90 data}\label{malt90}


\begin{figure*}
\begin{center}
\includegraphics[width=1.0\textwidth]{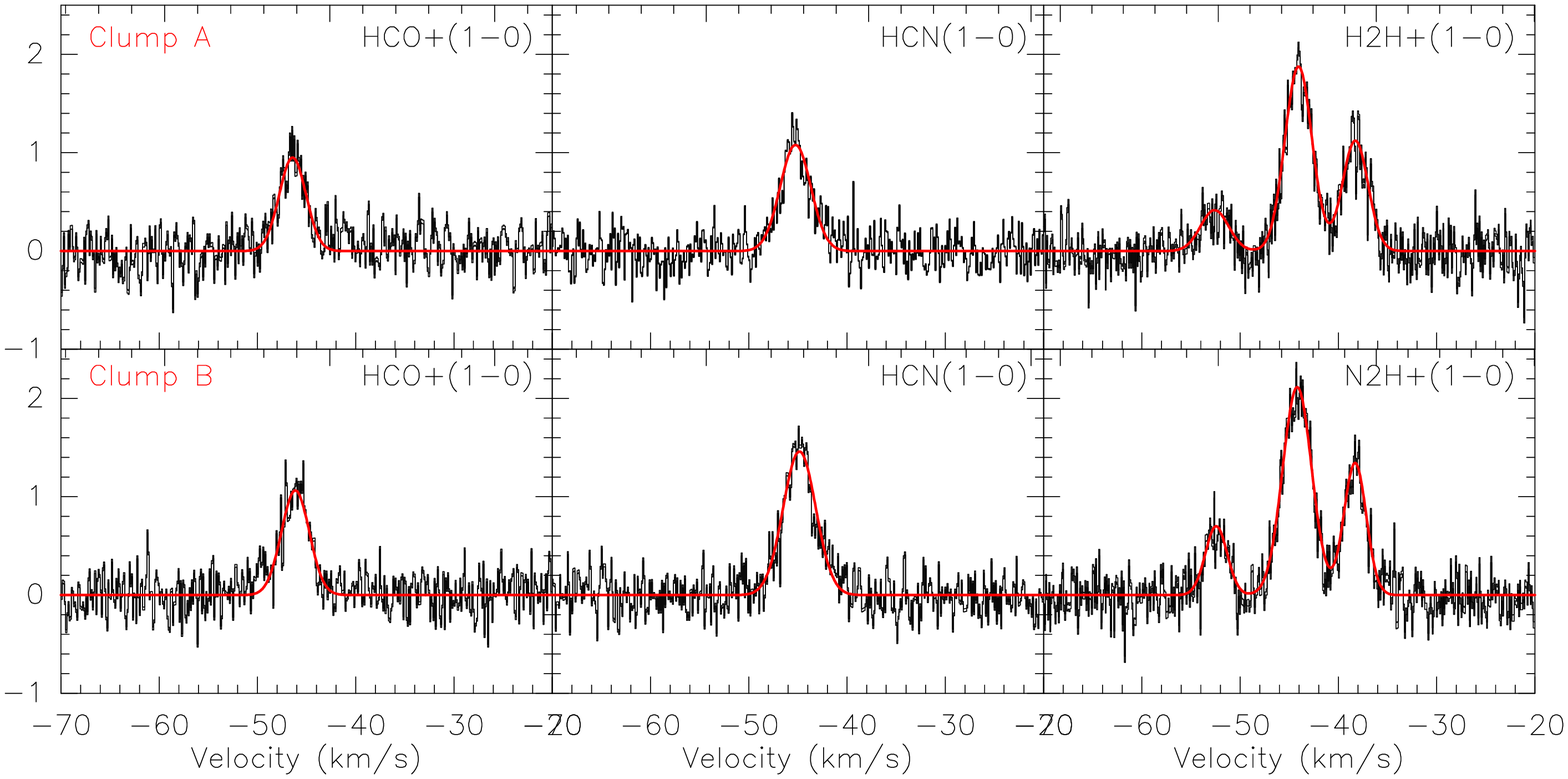}
\end{center}
\caption{HCO$^+$(1--0), HNC(1--0) and N$_2$H$^+$(1--0) profiles toward the \ca\ (top) and \cb\ (bottom).}
\label{malt90}
\end{figure*}


\begin{figure*}
\begin{center}
\includegraphics[width=12cm]{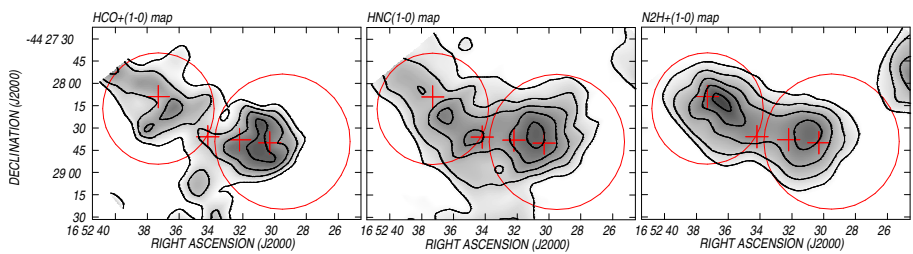}
\end{center}
\caption{HCO$^+$(1--0) (levs: 0.65, 0.7, 0.8 and 0.9 K),  HNC(1--0) (levs: 0.7, 0.85, 1.0 and 1.15 K) and N$_2$H$^+$(1--0) 
(levs: 0.8, 1.0, 1.2 and 1.4 K) maps. Red circles have radii equal to the effective radii of the clumps as seen in \tco, and are located at their  position. Crosses mark the location of the EGO candidates.}
\label{malt90mapas}
\end{figure*}


The MALT90 data cubes of this region show emission of HCO$^+$(1--0), HNC(1--0), and N$_2$H$^+$(1--0) molecules, which are the most often detected molecules towards IRDCs \citep{sanhueza2012,Rathborneetal2016}. These molecules have critical densities of 2$\times$10$^5$, 3$\times$10$^5$ and 3$\times$10$^5$ \cm, respectively \citep{sanhueza2012}, allowing us to infer a lower limit for the density of the clumps. These values are higher than the H$_2$ ambient densities derived from CO lines (see Table \ref{gas-para}). 

Spectra of these lines, averaged over the area of the clumps, are shown in Fig.~\ref{malt90mapas}.
By applying a gaussian fitting to the HCO$^+$ and HNC lines, a mean velocity of  --46 \kms\ was derived. For the case of the N$_2$H$^+$ line, a multi gaussian function was applied to take into account hyperfine structure. This molecule has seven hyperfine components, in two blended groups of three lines (groups 2 and 3) and one isolated component (group 1). The rest frequencies of groups 1, 2 and 3 are 93.176, 93.173, and 93.171 GHz, respectively. Its mean velocity is --44 \kms. The velocities of these lines coincide with the systemic velocities derived  from CO data.

These tracers of dense gas provide slightly different information: HCO$^+$ often shows infall signatures and outflow wings \citep[e.g.][]{Rawlings2004,Fuller2005}. However, in our case no signs of infall can be identified. 
HNC is specially preponderant in cold gas and is a commonly used as tracer of dense gas in molecular clouds. Finally, N$_2$H$^+$ is more resistant to freeze out on grains than the carbon-bearing species like CO. 

Figure~\ref{malt90mapas} shows the HCO$^+$(1--0), HNC(1--0) and N$_2$H$^+$(1--0) maps, integrated from --47.6 to --44.0 \kms, --46.7 to -43.1 \kms, and --47 to --41.4 \kms, respectively. Both clumps are detected in the three molecular lines indicating mean densities of 10$^5$ \cm. Carbon-bearing species, like CO, tend to disappear from the gas phase in the high density centers of the cores, while nitrogen-bearing species like N$_2$H$^+$ survive almost unaffected up to much higher densities. That is evidenced in the fact that the spacial distribution of the N$_2$H$^+$ emission is similar to dust continuum emission (see Figure~\ref{her-labo}).

Following \citet{Purcell2009} we estimated the N$_2$H$^+$ optical depth and column density. Assuming the line widths of the individual hyperfine components are all equal, the integrated intensities of the three blended groups should be in the ratio of 1:5:3 under optically thin conditions. The optical depth can then be derived from the ratio of the integrated intensities of any group using the following equation:
\begin{equation}
 \frac{\int T_{\rm mb,1} d{\rm v}}{\int T_{\rm mb,2} d{\rm v}} = \frac{1-exp(-\tau_1)}{1-exp(-\tau_2)} = \frac{1-exp(-\tau_1)}{1-exp(-a \tau_1)}
\end{equation}
\noindent where $a$ is the expected ratio of $\tau_1/\tau_2$ under optically thin conditions. 

To determine the optical depth we used only the intensity ratio of group 1/group 2, as \citet{Caselli1995} report anomalous excitation of the F$_1$,F = 1,0 $\rightarrow$ 1,1 and 1,2 $\rightarrow$ 1,2 components (in group 3). Thus we obtain $\tau_1$ = 0.11 and 0.19 for \ca\ and \cb, respectively.

Based on the expression for $T_{\rm mb}$ given by \citet*{RohlfsandWilson2004},  
the excitation temperature for N$_2$H$^+$ can be calculated with the following formula:
\begin{equation}
 T_{\rm exc} = 4.47 / ln \left[ 1 + \left(   \frac{T_{\rm mb}}{4.47( 1-exp(-\tau) )}  + 0.236   \right)^{-1} \right]
\end{equation}
\noindent and the column densities can be derived using \citep{Chen2013}:
\begin{eqnarray}
N({\rm N_2H^+}) = \frac{3kW}{8 \pi^3 \nu S \mu^2} \left( \frac{T_{\rm exc}}{T_{\rm exc}-T_{\rm bg}} \right) \left( \frac{\tau}{1 - exp(-\tau)} \right) \nonumber \\
\nonumber \\
\times Q(T_{\rm exc}) exp(E_u / k T_{\rm exc})
\end{eqnarray}
\noindent where $k$ is the Boltzmann constant, $W$ is the observed line integrated intensity (obtained from gaussian fit), $\nu$ is the frequency of the transition, 
and $S\mu^2$ is the product of the total torsion-rotational line strength and the square of the electric dipole moment. $T_{\rm exc}$ and $T_{\rm bg}$ are the excitation 
temperature and background brightness temperature, respectively. $E_u / k$ is the upper level energy in K, $Q(T_{\rm exc})$  is the partition function at temperature 
$T_{\rm exc}$ and $\tau$ is the optical depth. For group 1 the values of $\nu$, $S\mu^2$ and $E_u / k$ are 93176.2526\,MHz, 12.42\,D$^2$ and 4.47\,K, respectively. 
These values were taken from the SPLATALOGUE catalogue \footnote{SPLATALOGUE at http://www.splatalogue.net/}.

We derive $T_{\rm exc}$ = 17.4 and 12.5\,K and $N({\rm N_2H^+})$ = 8.7$\times10^{13}$ and 8.1$\times10^{13}$ for \ca\ and \cb, respectively. Considering an abundance
[N$_2$H$^+$]/[H$_2$] = 5$\times$10$^{-10}$ for dark molecular clouds \citep{ohishi1992} we obtain H$_2$ column densities of 1.7 and 1.6 $\times$10$^{23}$ for \ca\ and \cb, 
respectively. These column densities are higher than those obtained from CO calculations. Ambient densities are $n_{H_2} \approx$ 1.7$\times$10$^5$\,\cm\ for each clump
with an effective radii in  N$_2$H$^+$ of 0.53\,pc. This difference may be explained by considering that  the nitrogen-bearing 
species survive almost unaffected up to much higher densities that carbon-bearing molecules.

 \section{Warm and cold dust distribution}\label{dust}
 
\begin{figure*}
\centering
\includegraphics[angle=0,width=12cm]{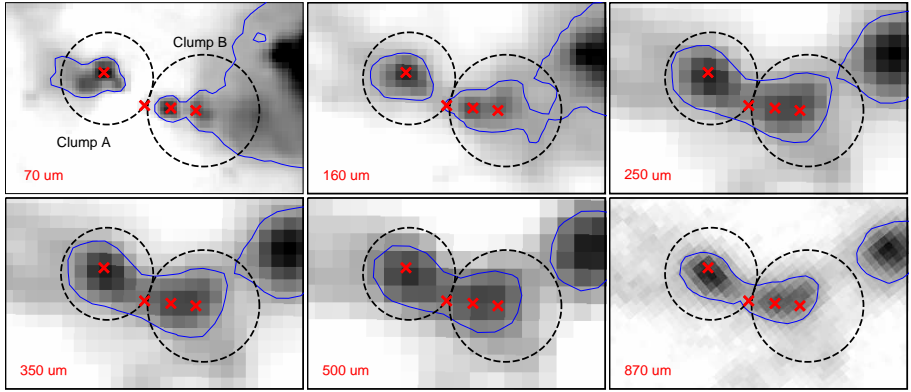}
\caption{FIR view of \sdc\ from Herschel (70-500\,\mic) and ATLASGAL data. The blue circles have the same meaning than in Figure~\ref{malt90mapas} and show the areas where the flux 
densities were integrated. The crosses mark the position of the EGOs. Contours correspond to 20\,{\it rms}.}
\label{her-labo}
\end{figure*}


\begin{figure*}
 \includegraphics[width=0.5\textwidth]{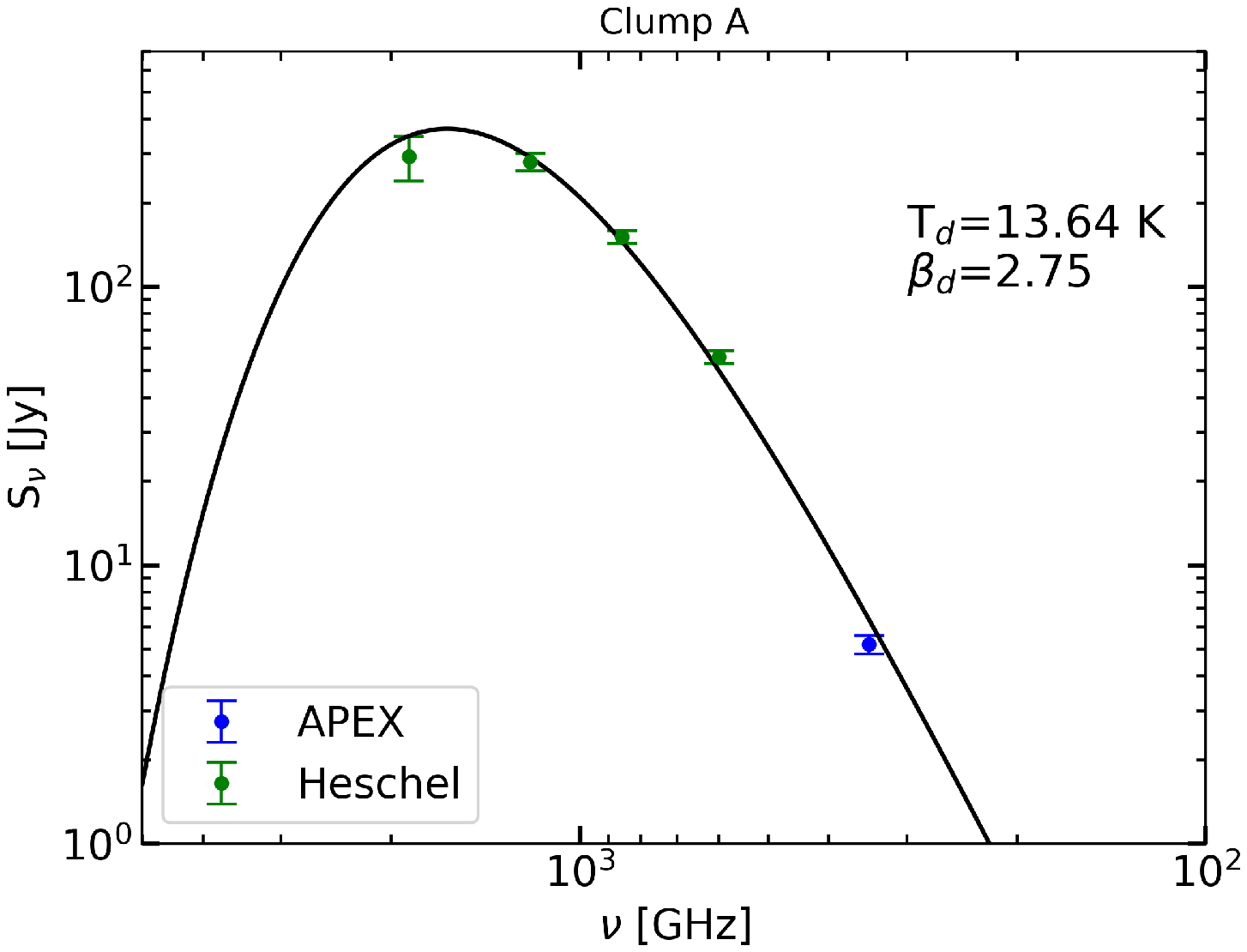}
 \includegraphics[width=0.5\textwidth]{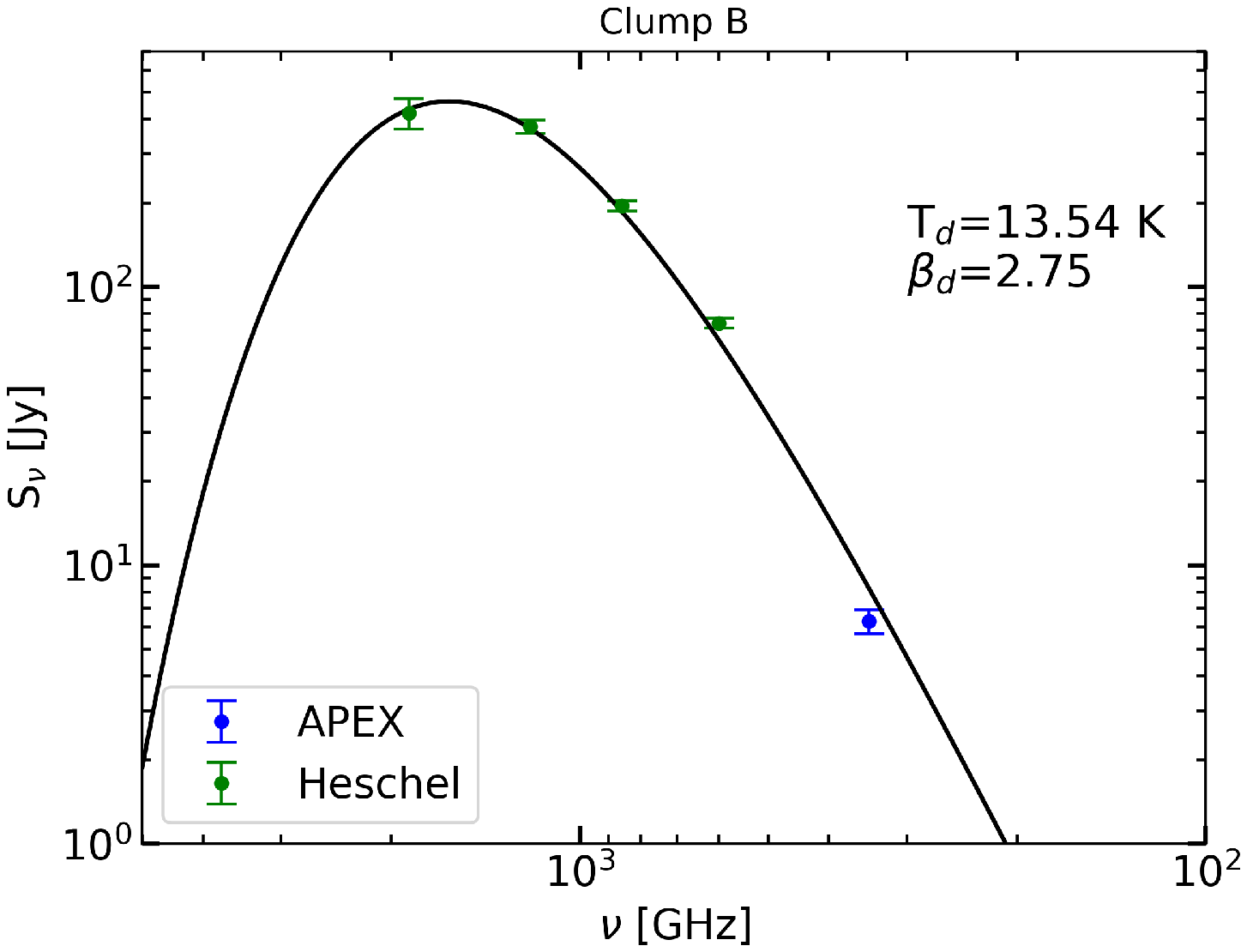}
 \caption{Spectral energy distributions (SEDs) for \ca\ (left) and \cb\ (right), obtained  from the fluxes at 160, 250, 350, and 500\,$\mu$m from Herschel, and 870\,$\mu$m from ATLASGAL.}
 \label{SEDs-clumps}
\end{figure*}

\begin{table*}
 \begin{changemargin}{-1cm}{-1cm}
\caption{Measured fluxes and derived parameters of the FIR clumps.}
\label{flujosIR}
\begin{tabular}{lccccccccc}
\hline
 & S$_{160}$  &S$_{250}$ & S$_{350}$ &S$_{500}$  & S$_{870}$ & T$_d$ &   $M_{\rm dust+gas}$  & $n_{\rm H_2}$  \\

 &    [Jy]    &   [Jy]   &    [Jy]   &  [Jy]     &  [Jy]   & [K]   &  [$M_{\odot}$]       &    [10$^3$ cm$^{-3}$]    \\

\hline

\ca &  293$\pm$53  &  281$\pm$21  & 151$\pm$8   &  56$\pm$3  & 5.2$\pm$0.4 & 13.6$\pm$0.47 & 1536$\pm$444 &  1.5$\pm$0.7  \\
\cb &  421$\pm$53  &  376$\pm$21  & 195$\pm$8   &  74$\pm$3  & 6.3$\pm$0.6 & 13.5$\pm$0.47 & 2088$\pm$600 &  1.4$\pm$0.6   \\

\hline
\end{tabular} 
\end{changemargin}
\end{table*}

Figure~\ref{her-labo} shows {\it Herschel} and ATLASGAL maps, in original resolution, of the \sdc\ region. At $\lambda <$ 160 \mic\ the beam resolution allows to
identify the emission associated with EGOs 1, 2, and 4. 
Warm dust coincident with the EGOs is also revealed by the emission in 24\,$\mu$m, which can be seen in Figure~\ref{Fig1}. The presence of this emission allows us to classify 
the clumps as ``active'', according to the classification of \citet{chambers2009}. These authors proposed an evolutionary sequence in which ``quiescent'' clumps (contains neither 
IR indicator) evolve into ``intermediate'' (contain either a ``green fuzzy'' or a 24 \mic\ point source, but not both), ``active''(caracterized  by the presence of a ``green fuzzy'' 
coincident with an embedded 24\,$\mu$m source, such as those observed toward EGOs 1, 2, and 4), and ``red'' clumps (dominated by 8 \mic\ emission, which  contains PAH features).
At $\lambda >$ 160\,$\mu$m two dust clumps are detected superimposed onto more extended submillimeter emission, indicating dust related to \ca\ and \cb. 

The characterization of the dust properties of these clumps is limited by the resolution of FIR data. We study the integrated dust properties of each clump from the
spectral energy distribution (SED) using the Herschel (160 up to 500 \mic) and ATLASGAL maps convolved at common beam resolution of the 500 \mic\ map (36\arcsec).
The flux densities are obtained from a circular aperture photometry integration (radio of 37\farcs6 and 45\farcs6 for \ca\ and \cb, respectively), and subtracting a
background level computed from a rectangular region (width = 52\farcs5 and height 25\farcs8 in the North of the clumps). For flux uncertainty estimation, we considered the
standard deviation of surface brightness in the background region and the flux calibration uncertainties. For each clumps, the final SED is depicted in Figure 7 and the
fluxes and their error bars are listed in Table 4. We perform a thermal dust fit to the data considering a single-component modified blackbody (grey-boby), which depend
on the optical depth, the emissivity spectral index ($\beta_d$) and the dust temperature (T$_d$). In the optically thin regime, the best fitting provides similar dust temperature and emissivity for both
clumps, with values of T$_d$ = 13.5$\pm$0.47 k and $\beta_d$ = 2.7. These dust temperatures are of the order on those found by \citet{Guzman2015} (18.6$\pm$0.2) for proto-stellar clumps.

For \ca\ and \cb, the total mass ($M_{\rm dust+gas}$) were calculated from the optical depth of the dust obtained from the fit, using the following expression 
 
\begin{equation}
 M_{dust+gas} = \frac{\tau_{\nu_{fit}}}{\kappa_{870}}\left(\frac{345 GHz}{\nu_{fit}}\right)^{\beta_d}d^2 R_{\rm d}
\end{equation}
 
\noindent where $\tau_{\nu_{fit}}$ is the dust optical depth (= 10$^{-26}$ $\times$ fit amplitude), $\nu_{fit}$=1\,Hz is the frequency of the fit, 
$R_{\rm d}$ = 100 is the typical gas-to-dust ratio, 
$d$ is the distance and $\kappa_{870}$=1.00 \cmdos\ gr$^{-1}$ is the dust opacity at 345\,GHz \citep*{ossenkopf1994}. 
The mass uncertainties are computed propagating the error bars for dust optical depth and distance.
Additional errors may result from taking different values for the gas-to-dust ratio (in our galaxy take typical values between 100 and 150).

The results are listed in Table~\ref{flujosIR}, 
which includes the derived masses of the clumps and their volume densities. Uncertainties in masses and ambient densities are about 35\% and 60\%, respectively. 
Total masses derived from molecular gas and dust emission are in good agrement (within errors).

\begin{figure*}
\centering
\includegraphics[angle=0,width=7cm]{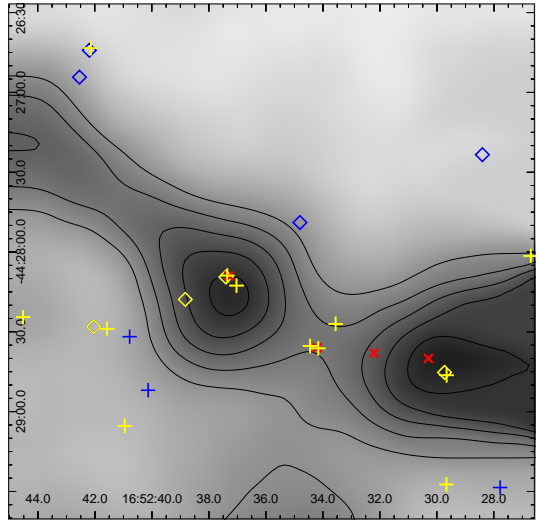}
\caption{Overlay of YSOs identified in the Spitzer and WISE catalogues and the molecular clumps in \tco. Contours have the same meaning as in Fig.~\ref{prom-13CO}. Yellow and blue symbols show the position of Class I  and Class II YSOs, while red crosses mark the location of EGOs. Plus signs correspond to Spitzer sources and diamonds to WISE ones.}
\label{ysos}
\end{figure*}


\section{Star formation}\label{dis}

\subsection{Search for additional YSOs coincident with the IRDC}

To investigate the coincidence of the IRDC with other candidate young stellar objects (YSOs), we analyze the characteristics of the point sources in the WISE 
and Spitzer catalogues \citep{Wright_2010, benjamin03} projected in the region. In Figure~\ref{ysos} we marked the positions of Class I and II YSO candidates identified in the area of the molecular clumps. 
An inspection of the figure reveals that EGOs 1, 3, and 4 coincide with identified YSOs.

\begin{figure*}
\begin{center}
\includegraphics[width=1.0\textwidth]{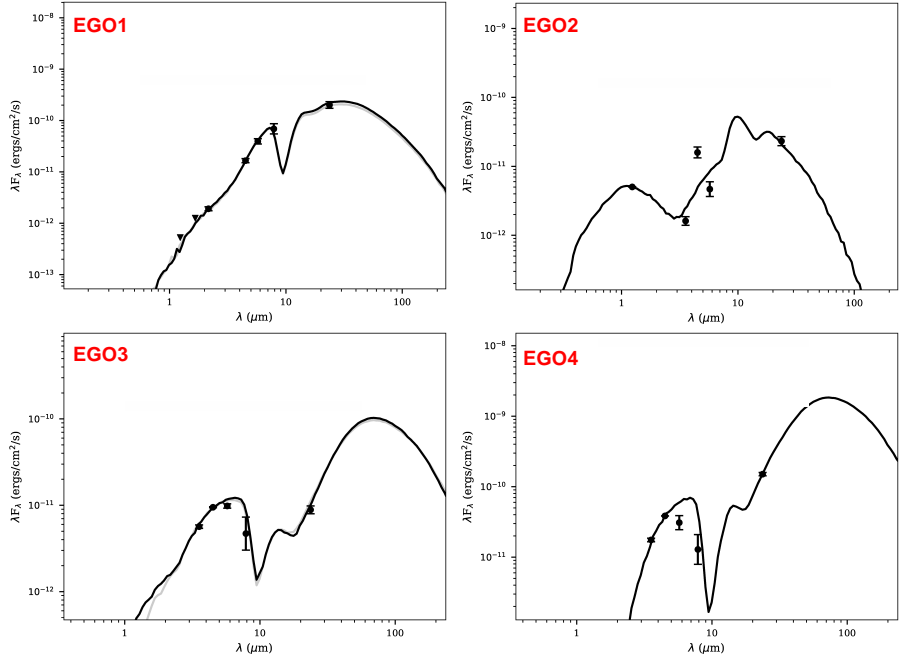}
\end{center}
\caption{Spectral energy distribution for the four candidate EGOs. }
\label{SEDs}
\end{figure*}

To study in some detail the nature of the EGOs we plotted their spectral energy distributions (SEDs) using the Robitaille's SED fitting tool \citep{robitaille2007}. 
To perform the analysis we were able to used available data from 2MASS, IRAC-GLIMPSE, and  MIPSGAL at 24\,$\mu$m only. 
The obtained SEDs are shown in Figure~\ref{SEDs}. Clearly, EGO 1, 3 and 4 have characteristics of YSOs, and as seen in Fig.~\ref{ysos} coincide with additional  star 
formation evidences. As regards EGO\,2, the fit is not good, it does not coincide with other sources or with the molecular clumps. So, its status is doubtful.    
In Table~\ref{param-sed} we display the main parameters obtained from the SEDs. Column 1 gives  the name of the source; columns 2 and 3,  the age and mass,
$M_{\rm stellar}$, of the central source; column 4,  the mass of the envelope, $M_{\rm env}$; column 5, the infall rate, $\dot{M}_{\rm acr}$; and column 6, the total luminosity. 
To perform the SEDs we adopted d = 3.6$\pm$0.5 kpc and a visual extinction of 3-4 mag. 

Following \citet{robitaille2007} an estimate of the evolutionary stage of the sources can be obtained based on the ratio of the infall rate and the mass of the central source. 
For the four candidate EGOs the ratio $\dot{M}_{\rm acr}/M_{\rm stellar} > $ 10$^{-6}$ indicates Stage 0/I. The age of some of the sources suggest that they are still immersed 
in their envelopes. According to the fitting, EGO\,4 would be the most massive object in the sample and  EGO\,2 seems to be the most evolved one. 
These results should be taken with caution because of the limitations of the fitting tool that may come from: (1)  YSOs are complex 3d objects with not very well axisymmetric 
density structures, so the models are incorrect compared to actual density distributions, (2) there are often mixtures of sources and many objects appear as a single YSO, but 
they are often two or more objects, and (3) even in the case of an isolated YSO variability is an additional complication, since the data with which SEDs are usually constructed 
belong to different elevations that have been performed in different years and can produce discontinuities in SEDs \citep{robitaille2008, deharveng2012, offner2012}.

As pointed out before, the emission at 24\,$\mu$m, detected toward EGO\,1, 3, and 4, suggests the existence of warm dust and embedded protostars \citep{jackson2008}.

\begin{table*}
 \begin{changemargin}{-2cm}{-2cm}
\centering
\caption[]{Physical parameters obtained from the SEDS for the EGOs}
\label{param-sed}
\begin{tabular}{lcccccc}
\hline
        &   Age    &    $M_{stellar}$  &    $M_{env}$     & $\dot{M}_{acr}$    &    $L$          & Stage \\
        &  10$^4$  &    M$_{\odot}$    &    M$_{\odot}$   & 10$^{-4}$\msunyr   & 10$^2L_{\odot}$ &           \\
\hline
 EGO\,1 &   0.13   &       2.1         &     0.7          &      0.1           &     1.3         &    0/I    \\
 EGO\,2 &   545    &       3.9         &   8\x10$^{-6}$   &      0             &     1.9         &    0/I    \\
 EGO\,3 &   15.7   &       4.1         &     7.1          &      1.4           &     0.8         &    0/I    \\
 EGO\,4 &   4.1    &       8.6         &     0.09         &      12.5          &     14.4        &    0/I    \\
\hline
\end{tabular}
\end{changemargin}
\end{table*}

\subsection{Evidence for inflow/outflow motions} \label{outflows}

\begin{figure*}
\begin{center}
\includegraphics[width=1.0\textwidth]{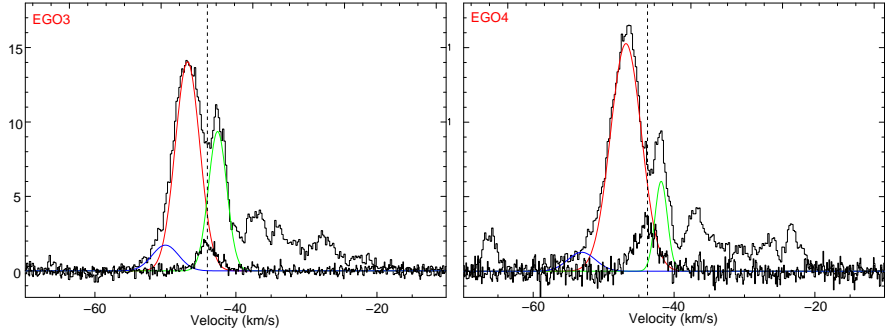}
\end{center}
\caption{\co\ and \cdo\ profiles in the direction to EGO\,3 (left) and EGO\,4 (right). In both pannels red and green lines correspond with gaussian fits from the \co\ double peak, 
and blue lines indicates a gaussian component fitting the possible blue outflow wings.}
\label{Fig6}
\end{figure*}


One way to visualize molecular outflows or infalls is to compare the optically thick \co\ emission line with the optically thin \cdo\ molecular line toward the candidate EGOs, 
as shown in Fig.~\ref{Fig6} for EGO\,3 and EGO\,4. The \co\ line present a double peak structure with the blue shifted peak brighter than the red-shifted one, and a minimum at 
the systemic velocity, while the \cdo\ line presents a simple peak centered at the absorption dip of the optically thick line. 
According to  \citet{chen2011} these spectra displays characteristics typical of infall motions: a double-peak structure with the blue-shifted peak brighter than the 
red-shifted one, while an emission peak of an optically thin line appears centered at the absorption dip of the optically thick line. According to these authors, the
infall motion is the only process that would consistently produce the blue profile asymmetry. The blue-shifted emission can be explained due the high-excitation 
approaching warm gas located on the far side of the center of contraction. The emission of this gas undergoes less extinction than the emission from the red-shifted 
receding nearside material, given that the excitation temperature of the molecules increases toward the center of the region \citep{zhou1992}. Thus, the spectra of
EGO\,3 and EGO\,4 suggest the presence of infall motions.

Following \citet{mardones1997}, we use the asymmetry parameter $\delta$v (the velocity difference between the peaks of the optically thick line and the optically thin lines, 
normalized by the FWHM of the thin line),
to quantify the blue asymmetry. The parameter is defined as \hbox{$\delta$v = (v$_{\rm thick}$ -- v$_{\rm thin}$) / $\Delta$v$_{\rm thin}$}.
An statistically significant excess of blue asymmetric line profiles with $\delta$v $<$ --0.25 
indicate that the molecular gas is infalling onto the clump. We consider the velocity and width of the \cdo, from Table \ref{tab:fits-lines}, as v$_{\rm thin}$ and 
$\Delta$v$_{\rm thin}$, and calculate v$_{\rm thick}$ from the gaussian fits of \co\ profiles (--46.9 and --45.3 from \ca\ and \cb, respectively). 
We find, for both \ca\ and \cb\, $\delta$v $\sim$ --0.06, which supports the infall hypothesis.

The interpretation presented above is supported by the $M_{\rm vir}/M({\rm H_2})$ ratio, previously derived for the objects. As the classical virial equilibrium analysis establishes, a 
ratio $M_{\rm vir}/M({\rm H_{2}}) \gtrsim$ 1 indicates that a clump has too much kinetic energy and is stable against gravitational collapse. Then, derived ratios suggest that \ca\ and \cb\ 
are not stable and could be collapsing.

For the sake of completeness, we calculate the eventual infall rate estimated from \citet*{Klaassen2007}
\begin{equation}
\label{infall_mass}
\dot{M}_{\rm inf}  = \frac{4}{3} \pi n_{H2} \mu m_H r_{\rm clump}^2 {\rm v}_{\rm inf}
\end{equation}

\noindent where $n_{\rm H2}$ is the H$_2$ volume density, $r_{\rm clump}$ is the linear radius of the infall region, and ${\rm v}_{\rm inf}$ is the infall velocity 
of the material. 
One way to estimate the infall velocity is considering the two layer radiative transfer model of \citet{myers1996}. 
Form the \co\ and \cdo\ spectra corresponding to \ca\ from Fig.~\ref{perfiles}, we calculated v$_{\rm inf}=$ 0.53 \kms using equation (9) in \citet{myers1996}.
In order to get a better estimate, we also calculated the infall velocity using the Hill5 model \citep{DeVriesandMyers2005} in the PySpecKit package \citep{GinsburgandMirocha2011}
\footnote{https://pyspeckit.readthedocs.io/en/latest/hill5infall\_model.html}. 
The Hill5 model employs an excitation temperature profile increasing linearly toward the center, rather than the two-slab model of \citet{myers1996}, 
so the Hill5 model is thought to provide a better fit of infall motions \citep{DeVriesandMyers2005}. From this model we obtained v$_{\rm inf}=$ 1.53$\pm$0.26 \kms.
Considering this value of v$_{\rm inf}=$ and the parameters listed in Table \ref{gas-para}, we obtained $\dot{M}_{\rm inf} \sim$ 7.47$\times$10$^{-4}$ \msunyr.
This result is consistent with those derived from surveys of massive YSOs such as \citet*{Klaassen2007}. 
\cb\ does not show clear observational evidences of collapse.

A typical outflow appears as spatially confined wings beyond the emission from the cloud core. For EGO\,3 and EGO\,4, we can distinguish the presence of  \co\ blue wings 
from $\sim$ --60 to --50 \kms, shown in the spectra fit in  Figure~\ref{Fig6}, while the velocity range of the optically thin \cdo\ emission is $\sim$ --47 to --40 \kms. 
The dashed line marks the systemic velocity of the clumps, that coincides with the central velocity of the \cdo\ line. 
Red wings can not be identified, although multiple components in the \co\ emission are present.


\begin{figure}
\begin{center}
\includegraphics[width=0.65\textwidth]{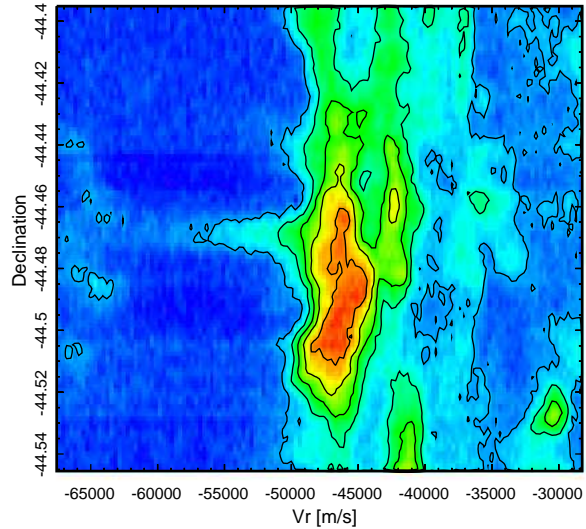}
\end{center}
\caption{Position-velocity diagram toward EGO\,4 in \co\ (Dec. vs. velocity for a fixed value of R.A. = 16:52:35). 
The color scale indicates T$_{mb}$ and declination is indicated in fractions of degrees.}
\label{Fig10}
\end{figure}


The typical signature of outflows can be seen in the position-velocity diagram. 
As shown in Fig.~\ref{Fig10} for EGO\,4, the extension of the emission toward velocities of about --55 \kms would correspond to the blue wings shown in Figure~\ref{Fig6}.
Note that the declination of this extension and the right ascension of the map coincides with the position of EGO\,4. 

These characteristics suggest that, although they are not spatially resolved, there may be outflows associated with EGO\,3 and EGO\,4.

\section{Summary}\label{summary}

Based on $^{12}$CO(2-1), $^{13}$CO(2-1), and C$^{18}$O(2-1) images obtained using the APEX telescope and  high density molecular tracers from the MALT90 survey 
such as HCO$^+$(1-0), HNC(1-0) and N$_2$H$^+$(1-0) we investigate the molecular component of the \irdc. 
The $^{13}$CO(2-1) and C$^{18}$O(2-1) along with MALT90 data reveal  two molecular clumps (\ca\ and \cb) 
linked to  \sdc\ with systemic velocities of --44.0\,\kms, indicating a kinematical distance of 3.6$\pm$0.5\,kpc.  Four EGOs (EGO\,1 at RA,Dec = 16:52:30.3, --44:28:40.0; 
EGO\,2 at RA,Dec = 16:52:32.2, --44:28:38.0; EGO\,3 at RA,Dec = 16:52:34.2, --44:28:36.0 and EGO\,4 at RA,Dec = 16:52:37.3, --44:28:09.9) 
coincide with the molecular clumps. 
We calculate masses of 2400$\pm$960 and 1200$\pm$480\,\msun, for \ca\ and \cb, respectively, and H$_2$ ambient densities $>$10$^4$\,\cm. Ambient densities estimated using the N$_2$H$^+$(1-0) line are higher ($\sim$ 10$^5$\,\cm) and agree with the critical density of this molecule, which  would be tracing the densest part of the clumps.

Both \ca\ and \cb\ are detected in the FIR (Herschel images) at $\lambda >$ 160\,$\mu$m. In the NIR and MIR, at $\lambda <$ 160\,$\mu$m (24 and 70\,$\mu$m), 
three of four EGOs seem to be resolved mainly at 24\,$\mu$m. Molecular masses derived from the emission at 870\,$\mu$m are roughly in agreement with those calculated from 
the molecular lines.

Spectral energy distributions (SEDs) for \ca\ and \cb\ built using fluxes in the FIR indicate dust temperatures of 13.5\,K, typical for an IRDC. 

Our search for additional signs of star formation indicates that some of the EGOs coincides with young stellar objects classified as Class I detected from point sources 
of the Spitzer and WISE catalogs. 

Additionally, the  \co\ spectra toward EGO\,3 and EGO\,4 present a double-peak structure with the blue-shifted peak brighter than the red-shifted one,
whith the maximum in the \tco\ and \cdo\ spectra coincident with the absorption dip, which disclose the existence of infall material. 
This fact together with the values obtained from $M_{\rm vir}/M({\rm H_2)}$ ratio derived for the clumps reveal that this would be collapsing.
Blue extended wings in the \cdo\ spectra are also present toward these EGOs suggesting the presence of outflows.

\section*{Acknowledgements}

This project was partially financed by CONICET of Argentina under projects PIP 00356, and PIP 00107 and from UNLP, projects PPID092, PPID/G002, and 11/G139. 
M.R. wishes to acknowledge support from CONICYT (CHILE) through FONDECYT grant No1140839.  
VF acknowledges support from CONICYT Astronomy Program-2015 Research Fellow GEMINI-CONICYT (32RF0002). V.F. also acknowledges support 
from the Faculty of the European Space Astronomy Centre (ESAC), and would like to thank Ivan Valtchanov, Bruno Altieri, and Luca Conversi 
for their support and valuable assistance in {\it Herschel} data processing. 
We thank to the referee for the careful reading of the manuscript and constructive comments which largely improved this presentation. 
This work is based [in part] on observations made with the Spitzer Space Telescope, which was operated by the Jet Propulsion Laboratory, 
California Institute of Technology  under a contract with NASA.

\bibliography{biblio}

\begin{thebibliography}
\expandafter\ifx\csname natexlab\endcsname\relax\def\natexlab#1{#1}\fi
\expandafter\ifx\csname href\endcsname\relax
  \def\href#1#2{}\fi
\expandafter\ifx\csname urllinklabel\endcsname\relax
  \def\urllinklabel{[LINK]}\fi
\expandafter\ifx\csname adsurllinklabel\endcsname\relax
  \def\adsurllinklabel{[ADS]}\fi

\bibitem[{{Allen}(1973)}]{allen1973}
{Allen}, C.~W. 1973, {Astrophysical quantities}


\bibitem[{{Battersby} {et~al.}(2010){Battersby}, {Bally}, {Jackson},
  {Ginsburg}, {Shirley}, {Schlingman}, \& et~al.}]{battersby2010}
{Battersby}, C., {Bally}, J., {Jackson}, J.~M., {Ginsburg}, A., {Shirley},
  Y.~L., {Schlingman}, W., \& et~al. 2010, \apj, 721, 222


\bibitem[{{Benjamin} {et~al.}(2003){Benjamin}, {Churchwell}, {Babler}, {Bania},
  {Clemens}, \& {Wolfire}}]{benjamin03}
{Benjamin}, R.~A., {Churchwell}, E., {Babler}, B.~L., {Bania}, T.~M.,
  {Clemens}, D.~P., \& {Wolfire}, M.~G. 2003, \pasp, 115, 953


\bibitem[{{Bergin} \& {Tafalla}(2007)}]{bergin2007a}
{Bergin}, E.~A. \& {Tafalla}, M. 2007, \araa, 45, 339


\bibitem[{{Beuther} {et~al.}(2002){Beuther}, {Schilke}, {Gueth}, {McCaughrean},
  {Andersen}, \& {Sridharan}}]{beuther2002}
{Beuther}, H., {Schilke}, P., {Gueth}, F., {McCaughrean}, M., {Andersen}, M.,
  \& {Sridharan}, T.~K. e.~a. 2002, \aap, 387, 931


\bibitem[{{Brand} \& {Blitz}(1993)}]{brand1993}
{Brand}, J. \& {Blitz}, L. 1993, \aap, 275, 67


\bibitem[{{Breen} {et~al.}(2013){Breen}, {Ellingsen}, {Contreras}, {Green},
  {Caswell}, {Stevens}, {Dawson}, \& {Voronkov}}]{Breenetal2013}
{Breen}, S.~L., {Ellingsen}, S.~P., {Contreras}, Y., {Green}, J.~A., {Caswell},
  J.~L., {Stevens}, J.~B., {Dawson}, J.~R., \& {Voronkov}, M.~A. 2013, \mnras,
  435, 524


\bibitem[{{Bronfman} {et~al.}(1996){Bronfman}, {Nyman}, \&
  {May}}]{bronfman1996}
{Bronfman}, L., {Nyman}, L.~A., \& {May}, J. 1996, A$\&$ASS, 115, 81


\bibitem[{{Burton} {et~al.}(2002){Burton}, {Walsh}, \&
  {Balasubramanyam}}]{Burtonetal2002}
{Burton}, M., {Walsh}, A., \& {Balasubramanyam}, R. 2002, in Astronomical
  Society of the Pacific Conference Series, Vol. 267, Hot Star Workshop III:
  The Earliest Phases of Massive Star Birth, ed. P.~{Crowther}, 355


\bibitem[{{Cappa} {et~al.}(2016){Cappa}, {Duronea}, {Firpo}, {Vasquez},
  {L{\'o}pez-Caraballo}, {Rubio}, \& et~al.}]{cappa2016}
{Cappa}, C.~E., {Duronea}, N., {Firpo}, V., {Vasquez}, J.,
  {L{\'o}pez-Caraballo}, C.~H., {Rubio}, M., \& et~al. 2016, \aap, 585, A30


\bibitem[{{Carey} {et~al.}(2005){Carey}, {Noriega-Crespo}, {Price}, {Padgett},
  {Kraemer}, {Indebetouw}, \& et~al.}]{carey2005}
{Carey}, S.~J., {Noriega-Crespo}, A., {Price}, S.~D., {Padgett}, D.~L.,
  {Kraemer}, K.~E., {Indebetouw}, R., \& et~al. 2005, in Bulletin of the
  American Astronomical Society, Vol.~37, American Astronomical Society Meeting
  Abstracts, 1252


\bibitem[{{Caselli} {et~al.}(1995){Caselli}, {Myers}, \&
  {Thaddeus}}]{Caselli1995}
{Caselli}, P., {Myers}, P.~C., \& {Thaddeus}, P. 1995, \apjl, 455, L77


\bibitem[{{Caswell} {et~al.}(2010){Caswell}, {Fuller}, {Green}, {Avison},
  {Breen}, {Brooks}, \& et~al.}]{caswell2010}
{Caswell}, J.~L., {Fuller}, G.~A., {Green}, J.~A., {Avison}, A., {Breen},
  S.~L., {Brooks}, K.~J., \& et~al. 2010, \mnras, 404, 1029


\bibitem[{{Chambers} {et~al.}(2009){Chambers}, {Jackson}, {Rathborne}, \&
  {Simon}}]{chambers2009}
{Chambers}, E.~T., {Jackson}, J.~M., {Rathborne}, J.~M., \& {Simon}, R. 2009,
  \apjs, 181, 360


\bibitem[{{Chen} {et~al.}(2011){Chen}, {Ellingsen}, {Shen}, {Titmarsh}, \&
  {Gan}}]{chen2011}
{Chen}, X., {Ellingsen}, S.~P., {Shen}, Z.-Q., {Titmarsh}, A., \& {Gan}, C.-G.
  2011, \apjs, 196, 9


\bibitem[{{Chen} {et~al.}(2013){Chen}, {Gan}, {Ellingsen}, {He}, {Shen}, \&
  {Titmarsh}}]{Chen2013}
{Chen}, X., {Gan}, C.-G., {Ellingsen}, S.~P., {He}, J.-H., {Shen}, Z.-Q., \&
  {Titmarsh}, A. 2013, \apjs, 206, 22


\bibitem[{{Chen} {et~al.}(2010){Chen}, {Shen}, {Li}, {Xu}, \&
  {He}}]{Chenetal2010}
{Chen}, X., {Shen}, Z.-Q., {Li}, J.-J., {Xu}, Y., \& {He}, J.-H. 2010, \apj,
  710, 150


\bibitem[{{Contreras} {et~al.}(2013){Contreras}, {Schuller}, {Urquhart},
  {Csengeri}, {Wyrowski}, {Beuther}, \& et~al.}]{contreras2013}
{Contreras}, Y., {Schuller}, F., {Urquhart}, J.~S., {Csengeri}, T., {Wyrowski},
  F., {Beuther}, H., \& et~al. 2013, \aap, 549, A45


\bibitem[{{Cyganowski} {et~al.}(2008){Cyganowski}, {Whitney}, {Holden},
  {Braden}, {Brogan}, {Churchwell}, \& et~al.}]{cyganowski2008}
{Cyganowski}, C.~J., {Whitney}, B.~A., {Holden}, E., {Braden}, E., {Brogan},
  C.~L., {Churchwell}, E., \& et~al. 2008, \aj, 136, 2391


\bibitem[{{De Vries} \& {Myers}(2005)}]{DeVriesandMyers2005}
{De Vries}, C.~H. \& {Myers}, P.~C. 2005, \apj, 620, 800


\bibitem[{{Deharveng} {et~al.}(2012){Deharveng}, {Zavagno}, {Anderson},
  {Motte}, {Abergel}, {Andr{\'e}}, \& et~al.}]{deharveng2012}
{Deharveng}, L., {Zavagno}, A., {Anderson}, L.~D., {Motte}, F., {Abergel}, A.,
  {Andr{\'e}}, P., \& et~al. 2012, \aap, 546, A74


\bibitem[{{Dickman}(1978)}]{dickman1978}
{Dickman}, R.~L. 1978, \apjs, 37, 407


\bibitem[{{Ellingsen}(2006)}]{Ellingsen2006}
{Ellingsen}, S.~P. 2006, \apj, 638, 241


\bibitem[{{Foster} {et~al.}(2011){Foster}, {Jackson}, {Barnes}, {Barris},
  {Brooks}, {Cunningham}, \& et~al.}]{Foster11}
{Foster}, J.~B., {Jackson}, J.~M., {Barnes}, P.~J., {Barris}, E., {Brooks}, K.,
  {Cunningham}, M., \& et~al. 2011, \apjs, 197, 25


\bibitem[{{Foster} {et~al.}(2013){Foster}, {Rathborne}, {Sanhueza},
  {Claysmith}, {Whitaker}, {Jackson}, \& et~al.}]{Foster13}
{Foster}, J.~B., {Rathborne}, J.~M., {Sanhueza}, P., {Claysmith}, C.,
  {Whitaker}, J.~S., {Jackson}, J.~M., \& et~al. 2013, PASA, 30, e038


\bibitem[{{Fuller} {et~al.}(2005){Fuller}, {Williams}, \&
  {Sridharan}}]{Fuller2005}
{Fuller}, G.~A., {Williams}, S.~J., \& {Sridharan}, T.~K. 2005, \aap, 442, 949


\bibitem[{{Ginsburg} \& {Mirocha}(2011)}]{GinsburgandMirocha2011}
{Ginsburg}, A. \& {Mirocha}, J. 2011, {PySpecKit: Python Spectroscopic Toolkit}


\bibitem[{{Griffin} {et~al.}(2010){Griffin}, {Abergel}, {Abreu}, {Ade},
  {Andr{\'e}}, {Augueres}, \& et~al.}]{Griffin2010}
{Griffin}, M.~J., {Abergel}, A., {Abreu}, A., {Ade}, P.~A.~R., {Andr{\'e}}, P.,
  {Augueres}, J.-L., \& et~al. 2010, \aap, 518, L3


\bibitem[{{Guzm{\'a}n} {et~al.}(2015){Guzm{\'a}n}, {Sanhueza}, {Contreras},
  {Smith}, {Jackson}, {Hoq}, \& et~al.}]{Guzman2015}
{Guzm{\'a}n}, A.~E., {Sanhueza}, P., {Contreras}, Y., {Smith}, H.~A.,
  {Jackson}, J.~M., {Hoq}, S., \& et~al. 2015, \apj, 815, 130


\bibitem[{{Hou} \& {Han}(2014)}]{hou2014}
{Hou}, L.~G. \& {Han}, J.~L. 2014, \aap, 569, A125


\bibitem[{{Jackson} {et~al.}(2008){Jackson}, {Chambers}, {Rathborne}, {Simon},
  \& {Zhang}}]{jackson2008}
{Jackson}, J.~M., {Chambers}, E.~T., {Rathborne}, J.~M., {Simon}, R., \&
  {Zhang}, Q. Astronomical Society of the Pacific Conference Series, Vol. 387,
  , Massive Star Formation: Observations Confront Theory, ed.
  H.~{Beuther}H.~{Linz} \& T.~{Henning}, 44


\bibitem[{{Jackson} {et~al.}(2013){Jackson}, {Rathborne}, {Foster}, {Whitaker},
  {Sanhueza}, {Claysmith}, \& et~al.}]{Jackson13}
{Jackson}, J.~M., {Rathborne}, J.~M., {Foster}, J.~B., {Whitaker}, J.~S.,
  {Sanhueza}, P., {Claysmith}, C., \& et~al. 2013, PASA, 30, e057


\bibitem[{{Klaassen} \& {Wilson}(2007)}]{Klaassen2007}
{Klaassen}, P.~D. \& {Wilson}, C.~D. 2007, \apj, 663, 1092


\bibitem[{{Ladd} {et~al.}(2005){Ladd}, {Purcell}, {Wong}, \&
  {Robertson}}]{Ladd2005}
{Ladd}, N., {Purcell}, C., {Wong}, T., \& {Robertson}, S. 2005, PASA, 22, 62


\bibitem[{{MacLaren} {et~al.}(1988){MacLaren}, {Richardson}, \&
  {Wolfendale}}]{macLaren1988}
{MacLaren}, I., {Richardson}, K.~M., \& {Wolfendale}, A.~W. 1988, \apj, 333,
  821


\bibitem[{{Mardones} {et~al.}(1997){Mardones}, {Myers}, {Tafalla}, {Wilner},
  {Bachiller}, \& {Garay}}]{mardones1997}
{Mardones}, D., {Myers}, P.~C., {Tafalla}, M., {Wilner}, D.~J., {Bachiller},
  R., \& {Garay}, G. 1997, \apj, 489, 719


\bibitem[{{Molinari} {et~al.}(2010){Molinari}, {Swinyard}, {Bally}, {Barlow},
  {Bernard}, {Martin}, \& et~al.}]{molinari2010}
{Molinari}, S., {Swinyard}, B., {Bally}, J., {Barlow}, M., {Bernard}, J.-P.,
  {Martin}, P., \& et~al. 2010, \pasp, 122, 314


\bibitem[{{Myers} {et~al.}(1996){Myers}, {Mardones}, {Tafalla}, {Williams}, \&
  {Wilner}}]{myers1996}
{Myers}, P.~C., {Mardones}, D., {Tafalla}, M., {Williams}, J.~P., \& {Wilner},
  D.~J. 1996, \apjl, 465, L133


\bibitem[{{Offner} {et~al.}(2012){Offner}, {Robitaille}, {Hansen}, {McKee}, \&
  {Klein}}]{offner2012}
{Offner}, S.~S.~R., {Robitaille}, T.~P., {Hansen}, C.~E., {McKee}, C.~F., \&
  {Klein}, R.~I. 2012, \apj, 753, 98


\bibitem[{{Ohishi} {et~al.}(1992){Ohishi}, {Irvine}, \& {Kaifu}}]{ohishi1992}
{Ohishi}, M., {Irvine}, W.~M., \& {Kaifu}, N. 1992, in IAU Symposium, Vol. 150,
  Astrochemistry of Cosmic Phenomena, ed. P.~D. {Singh}, 171


\bibitem[{{Ossenkopf} \& {Henning}(1994)}]{ossenkopf1994}
{Ossenkopf}, V. \& {Henning}, T. 1994, A$\&$A, 291, 943


\bibitem[{{Ott}(2010)}]{Ott2010}
{Ott}, S. Astronomical Society of the Pacific Conference Series, Vol. 434, ,
  Astronomical Data Analysis Software and Systems XIX, ed. Y.~{Mizumoto}K.-I.
  {Morita} \& M.~{Ohishi}, 139


\bibitem[{{Peretto} \& {Fuller}(2009)}]{peretto2009}
{Peretto}, N. \& {Fuller}, G.~A. 2009, \aap, 505, 405


\bibitem[{{Poglitsch} {et~al.}(2010){Poglitsch}, {Waelkens}, {Geis},
  {Feuchtgruber}, {Vandenbussche}, {Rodriguez}, \& et~al.}]{Poglitsch2010}
{Poglitsch}, A., {Waelkens}, C., {Geis}, N., {Feuchtgruber}, H.,
  {Vandenbussche}, B., {Rodriguez}, L., \& et~al. 2010, \aap, 518, L2


\bibitem[{{Purcell} {et~al.}(2009){Purcell}, {Longmore}, {Burton}, {Walsh},
  {Minier}, {Cunningham}, \& et~al.}]{Purcell2009}
{Purcell}, C.~R., {Longmore}, S.~N., {Burton}, M.~G., {Walsh}, A.~J., {Minier},
  V., {Cunningham}, M.~R., \& et~al. 2009, \mnras, 394, 323


\bibitem[{{Purcell} {et~al.}(2012){Purcell}, {Longmore}, {Walsh}, {Whiting},
  {Breen}, \& {Britton}}]{purcell2012}
{Purcell}, C.~R., {Longmore}, S.~N., {Walsh}, A.~J., {Whiting}, M.~T., {Breen},
  S.~L., \& {Britton}, T. e.~a. 2012, \mnras, 426, 1972


\bibitem[{{Rathborne} {et~al.}(2006){Rathborne}, {Jackson}, \&
  {Simon}}]{rathborne2006}
{Rathborne}, J.~M., {Jackson}, J.~M., \& {Simon}, R. 2006, \apj, 641, 389


\bibitem[{{Rathborne} {et~al.}(2008){Rathborne}, {Jackson}, {Zhang}, \&
  {Simon}}]{rathborne2008}
{Rathborne}, J.~M., {Jackson}, J.~M., {Zhang}, Q., \& {Simon}, R. 2008, \apj,
  689, 1141


\bibitem[{{Rathborne} {et~al.}(2007){Rathborne}, {Simon}, \&
  {Jackson}}]{rathborne2007}
{Rathborne}, J.~M., {Simon}, R., \& {Jackson}, J.~M. 2007, \apj, 662, 1082


\bibitem[{{Rathborne} {et~al.}(2016){Rathborne}, {Whitaker}, {Jackson},
  {Foster}, {Contreras}, {Stephens}, {Guzm{\'a}n}, {Longmore}, {Sanhueza},
  {Schuller}, {Wyrowski}, \& {Urquhart}}]{Rathborneetal2016}
{Rathborne}, J.~M., {Whitaker}, J.~S., {Jackson}, J.~M., {Foster}, J.~B.,
  {Contreras}, Y., {Stephens}, I.~W., {Guzm{\'a}n}, A.~E., {Longmore}, S.~N.,
  {Sanhueza}, P., {Schuller}, F., {Wyrowski}, F., \& {Urquhart}, J.~S. 2016,
  PASA, 33, e030


\bibitem[{{Rawlings} {et~al.}(2004){Rawlings}, {Redman}, {Keto}, \&
  {Williams}}]{Rawlings2004}
{Rawlings}, J.~M.~C., {Redman}, M.~P., {Keto}, E., \& {Williams}, D.~A. 2004,
  \mnras, 351, 1054


\bibitem[{{Ren} {et~al.}(2011){Ren}, {Liu}, {Wu}, \& {Li}}]{ren2011}
{Ren}, J.~Z., {Liu}, T., {Wu}, Y., \& {Li}, L. 2011, \mnras, 415, L49


\bibitem[{{Robitaille}(2008)}]{robitaille2008}
{Robitaille}, T.~P. Astronomical Society of the Pacific Conference Series, Vol.
  387, , Massive Star Formation: Observations Confront Theory, ed.
  H.~{Beuther}H.~{Linz} \& T.~{Henning}, 290


\bibitem[{{Robitaille} {et~al.}(2007){Robitaille}, {Whitney}, {Indebetouw}, \&
  {Wood}}]{robitaille2007}
{Robitaille}, T.~P., {Whitney}, B.~A., {Indebetouw}, R., \& {Wood}, K. 2007,
  \apjs, 169, 328


\bibitem[{{Rohlfs} \& {Wilson}(2004)}]{RohlfsandWilson2004}
{Rohlfs}, K. \& {Wilson}, T.~L. 2004, {Tools of radio astronomy}


\bibitem[{{Sanhueza} {et~al.}(2010){Sanhueza}, {Garay}, {Bronfman}, {Mardones},
  {May}, \& {Saito}}]{sanhueza2010}
{Sanhueza}, P., {Garay}, G., {Bronfman}, L., {Mardones}, D., {May}, J., \&
  {Saito}, M. 2010, \apj, 715, 18


\bibitem[{{Sanhueza} {et~al.}(2012){Sanhueza}, {Jackson}, {Foster}, {Garay},
  {Silva}, \& {Finn}}]{sanhueza2012}
{Sanhueza}, P., {Jackson}, J.~M., {Foster}, J.~B., {Garay}, G., {Silva}, A., \&
  {Finn}, S.~C. 2012, \apj, 756, 60


\bibitem[{{Schuller} {et~al.}(2009){Schuller}, {Menten}, {Contreras},
  {Wyrowski}, {Schilke}, {Bronfman}, \& et~al.}]{schuller2009}
{Schuller}, F., {Menten}, K.~M., {Contreras}, Y., {Wyrowski}, F., {Schilke},
  P., {Bronfman}, L., \& et~al. 2009, \aap, 504, 415


\bibitem[{{Scoville} {et~al.}(1986){Scoville}, {Sargent}, {Sanders},
  {Claussen}, {Masson}, {Lo}, \& et~al.}]{Scoville86}
{Scoville}, N.~Z., {Sargent}, A.~I., {Sanders}, D.~B., {Claussen}, M.~J.,
  {Masson}, C.~R., {Lo}, K.~Y., \& et~al. 1986, \apj, 303, 416


\bibitem[{{Sobolev} {et~al.}(2005){Sobolev}, {Ostrovskii}, {Kirsanova},
  {Shelemei}, {Voronkov}, \& {Malyshev}}]{Sobolevetal2005}
{Sobolev}, A.~M., {Ostrovskii}, A.~B., {Kirsanova}, M.~S., {Shelemei}, O.~V.,
  {Voronkov}, M.~A., \& {Malyshev}, A.~V. Massive Star Birth: A Crossroads of
  Astrophysics, ed. , R.~{Cesaroni}M.~{Felli}E.~{Churchwell} \& M.~{Walmsley},
  174--179


\bibitem[{{Vassilev} {et~al.}(2008){Vassilev}, {Meledin}, {Lapkin}, {Belitsky},
  {Nystr{\"o}m}, {Henke}, \& et~al.}]{vassilev2008}
{Vassilev}, V., {Meledin}, D., {Lapkin}, I., {Belitsky}, V., {Nystr{\"o}m}, O.,
  {Henke}, D., \& et~al. 2008, \aap, 490, 1157


\bibitem[{{Wang}(2006)}]{Wang2006}
{Wang}, J. 2006, {A First X-ray View of Infrared Dark Clouds, Precursors to
  Star Clusters}, XMM-Newton Proposal


\bibitem[{{Wright} {et~al.}(2010){Wright}, {Eisenhardt}, {Mainzer}, {Ressler},
  {Cutri}, {Jarrett}, \& et~al.}]{Wright_2010}
{Wright}, E.~L., {Eisenhardt}, P.~R.~M., {Mainzer}, A.~K., {Ressler}, M.~E.,
  {Cutri}, R.~M., {Jarrett}, T., \& et~al. 2010, AJ, 140, 1868


\bibitem[{{Yang} {et~al.}(2017){Yang}, {Xu}, {Chen}, {Ellingsen}, {Lu}, {Ju},
  \& et~al.}]{yang2017}
{Yang}, W., {Xu}, Y., {Chen}, X., {Ellingsen}, S.~P., {Lu}, D., {Ju}, B., \&
  et~al. 2017, \apjs, 231, 20


\bibitem[{{Zhang} {et~al.}(2007){Zhang}, {Sridharan}, {Hunter}, {Chen},
  {Beuther}, \& {Wyrowski}}]{Zhang2007}
{Zhang}, Q., {Sridharan}, T.~K., {Hunter}, T.~R., {Chen}, Y., {Beuther}, H., \&
  {Wyrowski}, F. 2007, \aap, 470, 269


\bibitem[{{Zhou}(1992)}]{zhou1992}
{Zhou}, S. 1992, \apj, 394, 204


\end{thebibliography}

\altaffiltext{1}{Instituto Argentino de Radioastronom{\'{\i}}a, CONICET, CCT-La Plata, C.C.5., 1894, Villa Elisa, Argentina.}
\altaffiltext{2}{Facultad de Ciencias Astron\'omicas y Geof{\'{\i}}sicas, Universidad Nacional de La Plata, Paseo del Bosque s/n, 1900 La Plata,  Argentina.}
\altaffiltext{3}{Departamento de Astronom{\'{\i}}a, Universidad de Chile, Casilla 36, Santiago de Chile, Chile.}
\altaffiltext{4}{Departamento de Astronom\'{\i}a y F\'{\i}sica, Universidad de la Serena, La Serena, Chile.}
\altaffiltext{5}{Gemini Obsetrvatory, Southern Operations Center, C/o AURA, Casilla 603, La Serena, Chile.}
\altaffiltext{6}{Instituto de Astrof{\'{\i}}sica and Centro de Astro-Ingenier{\'{\i}}a, Facultad de F{\'{\i}}sica, Pontificia Universidad Cat\'olica de Chile, Av. Vicu\~na Mackenna 4860, 7820436 Macul, Santiago, Chile}

\end{document}